\documentclass[usenatbib]{mnras}

\usepackage{newtxtext,newtxmath}

\usepackage[T1]{fontenc}

\usepackage{graphicx}	
\usepackage{amsmath}	
\usepackage{float}
\usepackage{xcolor}
\usepackage{hyperref}
\numberwithin{equation}{section}

\newcommand\Ev[1]{\left\langle #1 \right\rangle}
\newcommand\Fb[1]{\left( #1 \right)}
\newcommand\Tb[1]{\left[ #1 \right]}

\renewcommand\vec[1]{\mathbfit{#1}} 

\title[Effects of AGN feedback]{Quantifying the impact of AGN feedback on the large-scale matter distribution using two- and three-point statistics}

\author[B. Saha \& S. Bose]{
Bipradeep Saha$^{1,2}$\thanks{E-mail: bisaha@mpia.de} and
Sownak Bose$^{3}$ \thanks{E-mail: sownak.bose@durham.ac.uk}
\\
$^{1}$Max-Planck-Institut f{\"u}r Astronomie, K{\"o}nigstuhl 17, D-69117 Heidelberg, Germany\\
$^{2}$Department of Physical Sciences, IISER - Kolkata, India\\
$^{3}$Institute for Computational Cosmology, Department of Physics, Durham University, South Road, Durham DH1 3LE, UK\\
}

\date{Accepted XXX. Received YYY; in original form ZZZ}

\pubyear{2024}

\begin{document}
\label{firstpage}
\pagerange{\pageref{firstpage}--\pageref{lastpage}}
\maketitle

\begin{abstract}
Feedback from active galactic nuclei (AGN) plays a critical role in shaping the matter distribution on scales comparable to and larger than individual galaxies. Upcoming surveys such as \textit{Euclid} and LSST aim to precisely quantify the matter distribution on cosmological scales, making a detailed understanding of AGN feedback effects essential. Hydrodynamical simulations provide an informative framework for studying these effects, in particular by allowing us to vary the parameters that determine the strength of these feedback processes and, consequently, to predict their corresponding impact on the large-scale matter distribution. We use the EAGLE simulations to explore how changes in subgrid viscosity and AGN heating temperature affect the matter distribution, quantified via  2- and 3-point correlation functions, as well as higher order cumulants of the matter distribution. We find that varying viscosity has a small impact ($\approx 10\%$) on scales larger than $1 h^{-1}$ Mpc, while changes to the AGN heating temperature lead to substantial differences, with up to 70\% variation in gas clustering on small scales ($\lesssim 1 h^{-1}$ Mpc). By examining the suppression of the power spectrum as a function of time, we identify the redshift range $z = 1.5 - 1$ as a key epoch where AGN feedback begins to dominate in these simulations. The 3-point function provides complementary insight to the more familiar 2-point statistics, and shows more pronounced variations between models on the scale of individual haloes. On the other hand, we find that effects on even larger scales are largely comparable.
\end{abstract}




\section{Introduction}
Cosmological simulations have greatly improved our understanding of the physics of galaxy formation and are widely used to guide the interpretation of observations and the design of new observational campaigns and instruments. Simulations are useful for testing how different physical processes affect galaxy formation, which can then be compared with  observations in order to develop a more complete understanding of the Universe. 

In the following decades, ambitious observational campaigns will aim to pin down the source of the accelerated expansion of the Universe -- whether it is a cosmological constant, an exotic source of dark energy, or resulting from modifications to General Relativity -- as well as the impact of the neutrinos in the Universe, and detailed statistics of primordial fluctuations to test for deviations from Gaussianity. To tackle these challenges, the scientific community will perform far more precise surveys of the Universe on large scales. The next generation of galaxy surveys will rely on different proxies of matter in the Universe to answer the questions raised above. Previous and ongoing surveys like the  Sloan Digital Sky Survey \citep[SDSS,][]{huff2014seeing}, the Kilo Degree Survey \citep[KiDS,][]{de2013kilo}, and the Dark Energy Survey \citep[DES,][]{troxel2018dark} have demonstrated the enormous potential of this era of `precision cosmology'. Many experiments are planned to obtain better constraining power on the cosmological model, such as the \textit{Euclid} \citep{laureijs2011euclid} and Vera C. Rubin Observatory's Legacy Survey of Space and Time \citep[LSST,][]{ivezic2019lsst}. Before inferring cosmological parameters from these surveys, we need predictions of the theoretical matter power spectrum, $P(k)$, which quantifies the amount of statistical power in a given Fourier mode of the matter over-density field. Previous work suggests that the $P(k)$ needs to be modeled to a precision such that $\Delta P$ is constrained to being $\approx 1\%$ at $k_{\mathrm{max}}$ = $10 h/\mathrm{Mpc}$ or even larger scales \citep[e.g.][]{HUTERER2005369, Hearin_2012}. 

While it is possible to model the total matter power spectra via (semi-)analytical methods \citep[see for eg.][]{Somerville1999, Brennan2019} or by using dark matter-only simulations, it is now increasingly clear that the baryonic process could significantly impact the distribution of matter, and that these effects need to be considered while analyzing the data from the next-generation surveys \citep{Daalen_2011, Hellwing2016, Chisari2018}. Thus a crucial improvement is required in modeling the large-scale structure, i.e., the description of the impact of galaxy formation on the distribution of matter. Processes that heat and cool the gas, re-distribute it or transform it into stars have to be included in the models. The main effect to include is the suppression of power at the scales of a few Mpc, which is associated with the gas ejected by Active Galactic Nuclei (AGN). A range of ``effective'' models  to account for this problem have been suggested \citep[e.g.][]{Mead_2015, Schneider_2015}. However, state-of-the art hydrodynamical simulations, which model the co-evolution of the dark and ordinary matter self-consistently, are the most accurate methods to model the baryonic effects. \citep[e.g.][]{vogelsberger2014properties, Springel_2017}. The success of these techniques depends on the flexibility of these models to capture the true underlying matter distribution. The most significant limiting factor for hydrodynamical simulations is that these they are computationally very expensive compared to the dark matter-only simulations and are more difficult to control, as the multitude of parameters used to describe baryonic processes and feedback are poorly constrained.

In this work, we investigate how varying the sub-grid physics for AGN affects the distribution of gas and thus the total matter power spectra. We use the Virgo Consortium's EAGLE Project \citep[Evolution and Assembly of GaLaxies and their Environments,][]{Schaye_2015}, which is a suite of cosmological smooth particle hydrodynamic simulations of the $\Lambda$ Cold Dark Matter ($\Lambda$CDM) Universe. The main models were run in volumes of $25-100$ co-moving Mpc (cMpc) and the resolution is sufficient enough to marginally resolve the Jeans scales in the warm $(T \sim 10^4~K)$ interstellar medium (ISM). The aim of the present study is to help us understand how the physics of the subgrid manifest the large scale matter distribution and identify the statistics that are most responsive to these changes, at least qualitatively.  

In order to assess the effect of AGN feedback on the matter distribution, we make use of the two-point correlation function (2pCF) and the power spectra of different matter components in a given simulation and compare them for different variations in the sub-grid model. We then compute the cross-correlation between the gas field and the black-hole field weighted by the instantaneous accretion rate to quantify the specific impact of the growth and evolution of black holes specifically. This enables us to constrain the both the spatial and temporal scales where AGN feedback start to impact the large-scale matter distribution. To quantify this distribution further, we consider the cumulants of the matter over-density field. Finally, to quantify the non-Gaussian nature of the overdensity field and the baryonic impacts on it, we investigate the three-point correlation function (3pCF) of the gaseous component of the matter distribution.

This manuscript is organized as follows. Section~\ref{sec:simulation} describes the simulations and the model variations that we use. In Section~\ref{sec:methods}, we describe the methods that we use for our analysis. In Section~\ref{sec:results}, we present our main results and explain the different features we observe, followed by a discussion and a summary of our main conclusions in Sections~\ref{sec:discussion} and~\ref{sec:conclusion}, respectively.

\section{Simulation details}
\label{sec:simulation}

For our primary dataset, we use the Virgo Consortium's EAGLE project. EAGLE makes use of the smooth particle hydrodynamics (SPH) method run with a modified version of the Tree-PM SPH code GADGET3, last described by \cite{Springel_2005}. The main modifications are the formulation of SPH, the time stepping, and, most importantly, the subgrid physics.

The simulation was calibrated to match the relation between stellar mass and halo mass, galaxies' present-day stellar mass function, and galaxy sizes \citep{Schaye_2015,Crain_2015}.
The subgrid physics used in EAGLE is based on that developed for OWLS \citep{Schaye_2010} and COSMO-OWLS \citep{Brun_2014}. It includes element-by-element radiative cooling for 11 elements, star formation, stellar mass-loss, energy feedback from star formation, gas accretion onto and mergers of supermassive BHs (black holes), and AGN feedback. However, there are several changes from OWLS; the most important ones are implementations of energy feedback from star formation (which is thermal rather than kinetic), the accretion of gas onto BHs (which accounts for angular momentum), and the star formation law (which depends on metallicity). As they are of primary importance to our present work, we describe the AGN feedback model employed in EAGLE in the following subsections.

\subsection{Gas accretion onto black holes}\label{sec:accretion}
The rate at which BHs accrete gas depends on the mass of the BH, the local density and temperature of the gas, the velocity of the BH relative to the ambient gas, and the angular momentum of the gas with respect to the BH. Specifically, the gas accretion rate, $\dot{m}_{\mathrm{accr}}$, is given by the minimum of the the Eddington rate:
\begin{equation}
    \dot{m_{\mathrm{Edd}}}=\frac{4\pi G m_{\mathrm{BH}} m_p}{\epsilon_r \sigma_T c}
\end{equation}
and
\begin{equation}
    \dot{m}_{\mathrm{accr}} = \dot{m}_{\mathrm{Bondi}}\times \min\Fb{C_{\mathrm{Visc}}^{-1}\Fb{c_s/V_\phi}^3, 1} ,\label{eq:accr}
\end{equation}
where $\dot{m}_{\mathrm{Bondi}}$ is the Bondi \& Hoyle rate for spherically symmetric accretion \citep{Bondi_1944}. It is given by :
\begin{equation}
    \dot{m}_{\mathrm{Bondi}} = \frac{4 \pi G^2 m_{BH}^2 \rho}{\Fb{c_s^2 + v^2}^{3/2}}.
\end{equation}
The mass growth rate of the BH is given by
\begin{equation}
    \dot{m}_{\mathrm{BH}} = \Fb{1 - \epsilon_r}\dot{m}_{\mathrm{accr}}.
\end{equation}
In the above equations, $m_p$ is the proton mass, $\sigma_T$ is the Thomson cross-section, $c$ is the speed of light, $\epsilon_r = 0.1$ is the radiative efficiency, $v$ is the relative velocity of the BH and the gas and finally, $V_\phi$ is the rotation speed of the gas around the BH computed using Equation 16 of \cite{Rosas_Guevara_2015}. Here, $C_{\mathrm{Visc}}$ is a free parameter related to the viscosity of a notional subgrid accretion disc. The factor $\Fb{c_s/V_\phi}^3/C_{\mathrm{Visc}}$ by which the Bondi rate is multiplied in Equation~\ref{eq:accr} is equivalent to the ratio of the Bondi and viscous time scales \citep[see][]{Rosas_Guevara_2015}. The critical ratio of $V_\phi/c_s$ above which angular momentum is assumed to suppress the accretion rate scales as $C_{\mathrm{Visc}}^{-1/3}$.  Thus, larger values of $C_{\mathrm{Visc}}$ correspond to a lower subgrid kinetic viscosity, and so act to \textit{delay} the growth of BHs by gas accretion and, by extension, the onset of quenching by AGN feedback.

\subsection{AGN feedback}\label{sec:AGN_feedback}
AGN feedback in the EAGLE simulation is implemented thermally and stochastically. By making the feedback stochastic, one can control the amount of energy per feedback event even if the mean energy injected per unit mass is fixed. The energy injection rate is given by $\epsilon_f\epsilon_e \dot{m}_{\mathrm{accr}}c^2$, where $\epsilon_f$ is the fraction of the radiated energy that couples with the interstellar medium (ISM). The value of $\epsilon_f$ needs to be chosen by calibrating to the observation. In \cite{Schaye_2015} it is justified that $\epsilon_f=0.15$ and $\epsilon_r=0.1$ is a suitable choice of these parameters.

Each BH carries a reservoir of feedback energy, $E_{\mathrm{BH}}$. After each time step $\Delta t$, energy equivalent to $\epsilon_f\epsilon_r\dot{m}_{\mathrm{accr}}c^2\Delta t$
is added to the reservoir.  Once a BH has stored sufficient energy to heat at least one fluid element of mass $m_\mathrm{g}$, the BH is allowed to the heat each of its SPH neighbours by a temperature $\Delta T_{\mathrm{AGN}}$, stochastically. For each neighbour, the heating probability is given by:
\begin{equation}
    P = \frac{E_{\mathrm{BH}}}{\Delta \epsilon_{\mathrm{AGN}}N_{\mathrm{ngb}}\Ev{m_\mathrm{g}}}
\end{equation}
where, $\epsilon_{\mathrm{AGN}}$ is the change in internal energy per unit mass corresponding to the temperature increment, $N_{\mathrm{ngb}}$ is the number of gas neighbours of the BH and $\Ev{m_g}$ is their mean mass.

Larger values of $\Delta T_{\mathrm{AGN}}$ yield more energetic feedback
events, generally resulting in reduced radiative losses \citep{Crain_2015}. However, larger values also make the feedback more intermittent. In general, the ambient density of gas local to the central BH of galaxies is greater than that of star-forming gas distributed throughout their discs, so a higher heating temperature is required to minimise numerical losses.

In this work, we used two different model variations from the EAGLE run:
\begin{itemize}
    \item {\tt ViscHi} and {\tt ViscLo}
    \item {\tt AGNdT8} and {\tt AGNdT9}
\end{itemize}
which correspond to variations in the subgrid physics parameters relevant to the processes described above. Their details are summarized in Table \ref{tb:variation}.
As mentioned above, these model variations affect gas accretion onto the black holes and the effective strength of AGN feedback, respectively \citep{Crain_2015, Schaye_2015}. Following the resolution scheme of \cite{Schaye_2015}, both physics variations and reference models are of intermediate resolution. This translates to an initial mass for gas particles  $m_{\rm gas} = 1.81\times 10^6 ~{\rm M_\odot}$ and the mass of dark matter particles of $m_{\rm DM} = 9.70\times 10^6 ~{\rm M_\odot}$.

\begin{table}
\centering
\begin{tabular}{ccccc} 
\hline\hline
Model Name & \begin{tabular}[c]{@{}c@{}}Box size\\ {[}$\mathrm{h^{-1}~Mpc}$]\end{tabular} & N    & $C_{\mathrm{Visc}}/2\pi$ & \begin{tabular}[c]{@{}c@{}}$\Delta T_{\mathrm{AGN}}$\\ $\log_{10} [k]$\end{tabular}  \\ 
\hline\hline
Ref        & 33.885                                                            & $752^3$ &  $10^0$             & $8.5$                                                                       \\
{\tt ViscHi}     & 33.885                                                             & $752^3$  & $\mathbf{10^{-2}}$       & $8.5$                                                                       \\
{\tt ViscLo}     & 33.885                                                             & $752^3$  & $\mathbf{10^2}$        & $8.5$                                                                       \\
{\tt AGNdT8}     & 33.885                                                             & $752^3$  & $10^0$        & $\mathbf{8}$                                                                         \\
{\tt AGNdT9}     & 33.885                                                             & $752^3$  & $10^0$        & $\mathbf{9}$                                                                         \\
{NoAGN} & 33.885 & $752^3$ &- &-\\
\hline
\end{tabular}
\caption{Table summarizing the parameter variation between different runs of the EAGLE simulation. Here $N$ is the total number of particles of each type (DM and Gas), $C_{\mathrm{Visc}}$ is the viscosity parameter and $\Delta T_{\mathrm{AGN}}$ is the AGN heating parameter. The bold quantities indicate the parameters that were  changed with respect to the reference values for that model variation. }
\label{tb:variation}
\end{table}

\section{Methods}
\label{sec:methods}
This section introduces the definition of various statistical tools used to characterise the distribution of the matter density field in our analysis.

If $\rho(\vec{x})$ is the matter density at the point $\vec{x}$, and $\bar{\rho}$ is the mean matter density of the universe, we define the density contrast as:
\[
    \delta\Fb{\vec{x}} = \frac{\rho\Fb{\vec{x}}}{\bar{\rho}} - 1.
\]
Then, the Fourier modes for the density contrast field for the set of $N$ particles with mass $m_i$ in a periodic box of length $L$ of volume $V_u$ is defined as \citep{Peebles_1980, mo2010galaxy}:
\begin{equation}
    \label{eq:fourier_density}
    \delta_{\vec{k}} = \frac{1}{M}\sum_{i=1}^N m_i\exp\Fb{-i\vec{x}\cdot\vec{k}} = \frac{1}{V_u}\int\delta\Fb{\vec{x}}\exp\Fb{-i\vec{x}\cdot\vec{k}}
\end{equation}
where $\sum m_i = M$ and the second equality is in the continuum limit.

\subsection{Spatial Correlation Functions}
$N$-point statistics are essential tools for quantifying a distribution of points in a field. In cosmology, correlation functions are used to quantify the clustering of objects in the Universe,  test hierarchical scenarios for structure formation, test Gaussianity of the initial conditions, and test various models for the clustering bias between luminous and dark matter.

The \textit{two-point correlation function} (2pCF) measures the excess probability of finding two correlated points separated by distance $r$ \citep{Peebles_1980}:
\begin{equation}
    \delta P = n^2\Tb{1 + \xi(r)}dV_1dV_2, 
\end{equation}
where $\delta P$ is the joint probability of finding particles in volume element $dV_1$ and $dV_2$ separated by distance $r$, $n$ is the mean number density  of tracers and $\xi(r)$ is the 2pCF. Its Fourier transform, known as the \textit{power spectrum}, $P(k)$, is given by:
\begin{align}
    P(k) &= V_u \Ev{|\delta_\vec{k}|^2} = V_u\Ev{\delta_{\vec{k}} \delta_{-\vec{k}}} \label{eq:pk}\\
    \xi\Fb{r} &= \int P(k) e^{-i\vec{k}\cdot \vec{x}} d^3\vec{x} = \frac{1}{2\pi^2}\int P(k)\frac{\sin\Fb{kx}}{kx}d^3\vec{x} \label{eq:xi}
\end{align}
where $\Ev{.}$ denotes the ensemble average, and $\delta_{\vec{k}}$ is as defined in Equation~\ref{eq:fourier_density}.

Similarly the \textit{three-point correlation function} (3pCF) measures the excess probability of finding three correlated points \citep[i.e., triangular configurations,][]{Peebles_1980}:
\begin{equation}
    \delta P = n^3\Tb{1 + \xi\Fb{r_a} + \xi\Fb{r_b} + \xi\Fb{r_c} + \zeta\Fb{r_a, r_b, r_c}}dV_1 dV_2 dV_3,
\end{equation}
where the terms inside $\Tb{.}$ is the full 3pCF and $\zeta\Fb{r_a, r_b, r_c}$ is the reduced 3pCF and $r_a, r_b, r_c$ are the three sides of the triangle formed by the three points. 

\begin{figure}
    \centering
    \includegraphics[width=7.4cm, height=7.9cm]{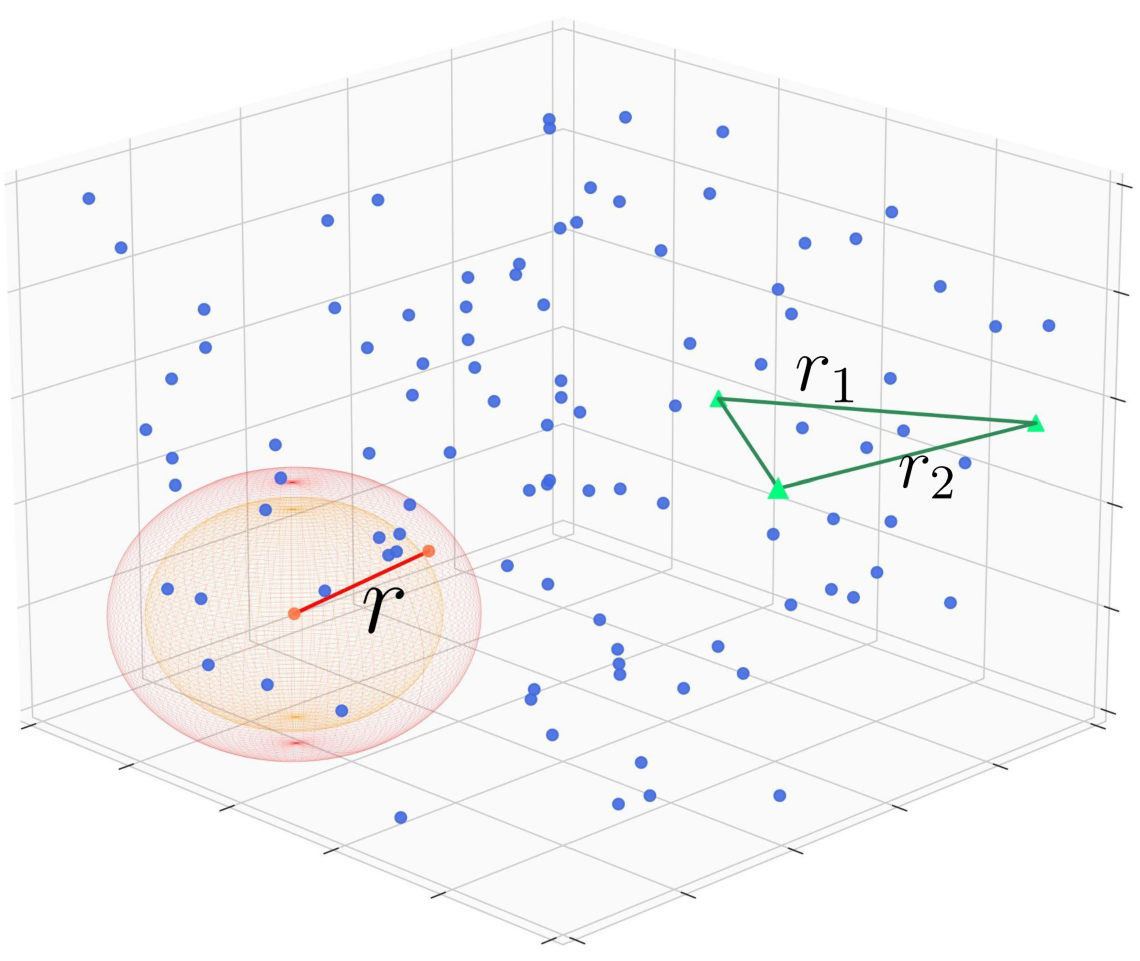}
    \caption{Illustration on how to measure 2pCF and 3pCF. For the 2pCF we sit at a point and compute the number of point inside a spherical shell of radius $r$, and thickness $dr$. We then spatially average this across the data to get the $DD$ (data-data) pair counts. We then repeat the same procedure for random data set and get $RR$ (random-random) pair counts. Similarly for 3pCF we count the triangles having side $r_1$ and $r_2$ in the observed and the random data set to get $DDD$ and $RRR$ value. Then using the generic estimators we can estimate the 2pCF and 3pCF.}
    \label{fig:fig1}
\end{figure}
Figure~\ref{fig:fig1} shows visual depictions of the 2- and 3-point correlation functions. Measuring the spatial correlation function from the above definitions is computationally expensive. Therefore we use estimators for computing each correlation function:
\begin{itemize}
    \item For computing the 2pCF and $P(k)$ we use the FFT based approach following the definitions in Equations~\ref{eq:pk} and~\ref{eq:xi}. In practice, we use the code by \cite{Pylians} for this purpose.
    \item For computing the 3pCF we use a Legendre polynomial decomposition described by \cite{Philcox2021}, which uses the generic estimator of the 3pCF:
    \begin{equation}\label{eq:3pcf}
        \zeta\Fb{r_a, r_b, r_c} = \frac{NNN}{RRR}
    \end{equation}
     as the primary definition, but uses Legendre polynomials to speed up the counting of number of triangles. Here, $N$ is defined as, $N:=D-R$. Hence, Equation~\ref{eq:3pcf} contains terms like $DDD, RRR, DRR, DDR$, where $DDD$ is number of triangles from the data set with sides $r_1 \pm \Delta r$  and $r_2 \pm \Delta r$, $RRR$ is the number of triangles from the  random data set with same side lengths. For $DRR$ we mix one part of the data with two parts of random data, and then compute the number of triangles with sides $r_1 \pm \Delta r$  and $r_2 \pm \Delta r$, and vice versa for $DDR$ counts.  The algorithm sits on each point in the dataset, and computes the spherical harmonic expansion of the density field in concentric spherical shells (radial bins) around that point, which is then combined to yield the multipole moments around this point, and then translation average is taken to yield, $\zeta_\ell$. Using this method, the 3pCF is given by:
    \begin{equation}\label{eq:3pCF}
        \zeta \Fb{\vec{r}_1, \vec{r}_2} = \sum_\ell \frac{\sqrt{2\ell + 1}}{4\pi}(-1)^\ell\zeta_\ell\Fb{r_1, r_2}L_\ell\Fb{\hat{\vec{r}}_1 \cdot \hat{\vec{r}}_2}
    \end{equation}
    where, $\Vec{r}_1$, $\Vec{r}_2$ parameterize the triangle, $\zeta_\ell$ are re-scaled Legendre Polynomials \citep{Slepian2015} and $L_\ell(x)$ are Legendre polynomials of order $\ell$. For our purpose, we set $\ell_{max}=5$, depending on the computational resources available and triangle configurations we are interested in. 
\end{itemize}

\subsection{Cumulants of Matter Distribution}\label{subsec:cumulants}
The non-linear evolution of the density field, $\delta$, drives the field and its distribution away from an initially Gaussian distribution \citep{bernardeau1992}. One way to study these deviations is by using cumulants or reduced moments \citep{Fry1984,Fry1985}. For a Gaussian field, the first two central moments are sufficient to characterize the full distribution. Thus, higher order cumulants of the density field are useful to characterize the non-linear density field in the presence of galaxy formation.

The $n-th$ cumulant of the density field, $\delta$, is defined by a recursive relation to the $n-$moment. It is expressed by the cumulant generating function, $K\Fb{\delta}$, as:
\begin{equation}
    \Ev{\delta^n} := \frac{\partial^n K\Fb{\delta}}{\partial t^n} = \frac{\partial^n \ln\Ev{e^{t\delta}}}{\partial t^n}.
\end{equation}
In our analysis we consider cumulants up to $5^{th}$ order. In terms of central moments, they are given by the following equations:
\begin{align*}
    \Ev{\delta^1}_c &= 0 \text{ (mean)}\\
    \Ev{\delta^2}_c &= \Ev{\delta^2} \equiv \sigma^2 \text{ (variance)}\\
    \Ev{\delta^3}_c &= \Ev{\delta^3} \text{ (skewness)}\\
    \Ev{\delta^4}_c &= \Ev{\delta^4} - 3\Ev{\delta^2}^2_c \text{ (kurtosis)}\\
    \Ev{\delta^5}_c &= \Ev{\delta^5} - 10\Ev{\delta^3}_c\Ev{\delta^2}_c,
\end{align*}
where the subscript $c$ denotes cumulants.

 Our analysis is based on the fact that the primordial matter density field is almost Gaussian. To remain unaffected by the impact of local maxima and minima in the density field, we first smooth the density field with a spherical filter of some size; this indeed degrades the resolution of the simulated density field to one that may be more readily calculated in observations. In our analysis, we also study how the cumulants change with smoothing scale chosen.

 Given that the box size of the simulations that we use in our study is very small, the quantities like 2pCF, 3pCF and Cumulants are unlikely to converge on the scales that can be probed. Thus, in order to circumvent this issue, rather than presenting absolute measurements of these quantities, we will use relative estimates -- either with respect to DM only simulations of the same box size, or among different models used in this study. As each of these variants are initialised from the same initial conditions, taking the ratio also helps beat down the effects of cosmic variance.

\section{Results}
\label{sec:results}
\subsection{Comparing different models}\label{subsec:models}

\begin{figure*}
    \centering
    \includegraphics[width=\textwidth]{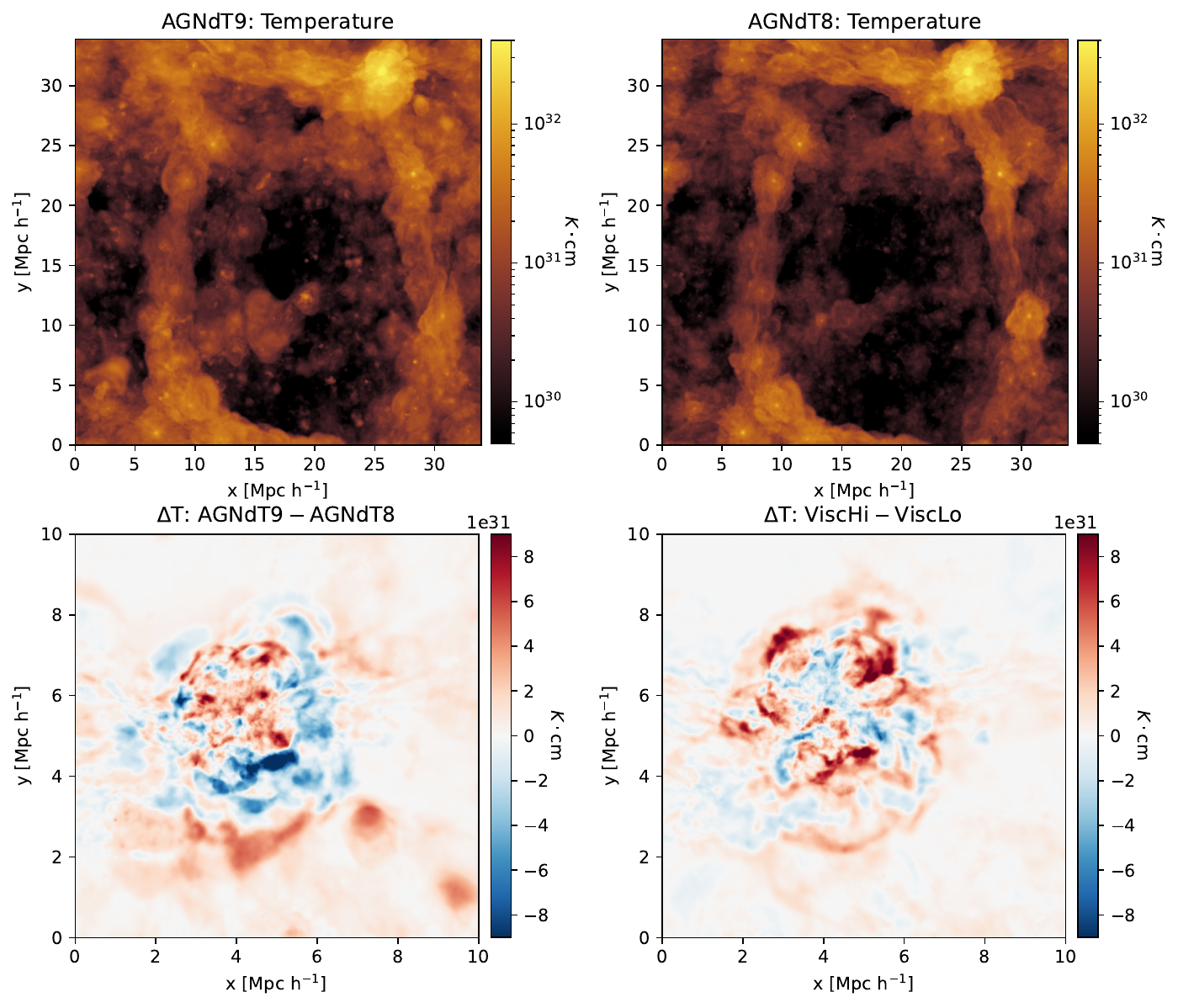}
    \caption{Temperature fields of the gaseous component across different models at $z=0$ projected along the $z-$direction. The top panel shows the temperature across the entire simulation box, while the bottom panel, shows the difference in temperature profile between two models in a 10 $\mathrm{Mpc~h^{-1}}$ region centred around the most massive halo. Gas in {\tt AGNdT9} is hotter than {\tt AGNdT8} due to higher heating temperature of BHs, while gas in {\tt ViscHi} is hotter than {\tt ViscLo} due to higher accretion rate which result in higher energy injection into the surrounding. The colour bar represents temperature integrated along the $z$-axis, and is in units of $K\cdot \mathrm{cm}$. }
    \label{fig:fig2}
\end{figure*}

\begin{figure}
    \centering
    \includegraphics[width=1\linewidth]{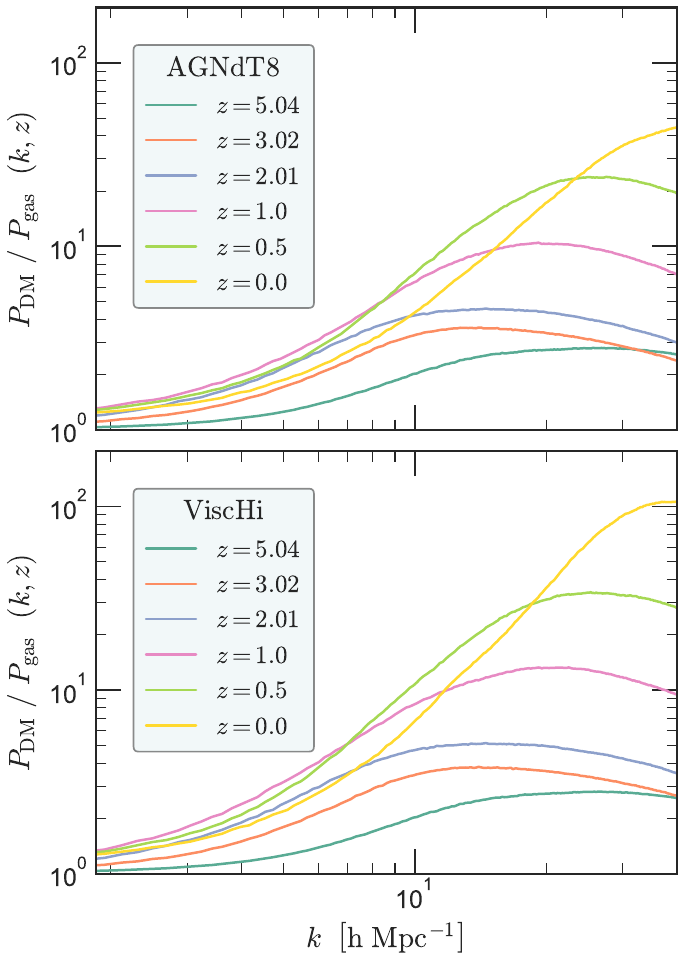}
    \caption{$P(k)$ at different epochs for {\tt AGNdT8} and {\tt ViscHi} model. With evolving time, clustering increases which boosts the $P(k)$. AGN feedback re-distributes matter, which suppresses the $P(k)$. The crossover events, which are departures from general trend of $P(k)$ with $z$, are indicative of the onset of dominance of AGN feedback. Similar behaviour is observed with {\tt AGNdT9} and {\tt ViscLo}, but with different intensity, which is studied later.}
    \label{fig:fig3}
\end{figure}

\begin{figure}
    \centering
    \includegraphics[width=1\linewidth]{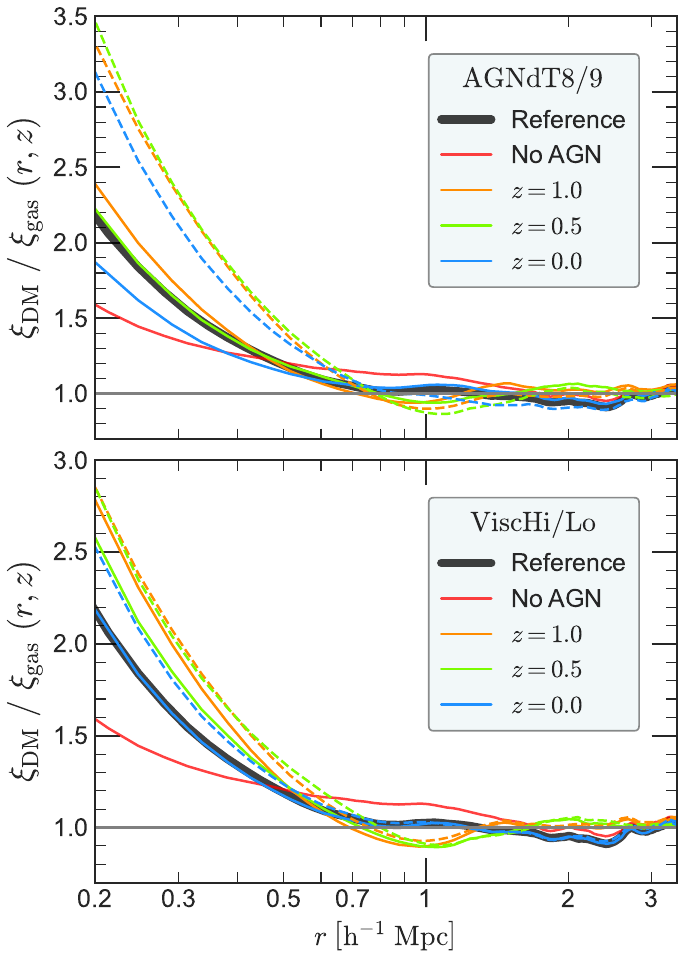}
    \caption{Comparing 2pCF of gas with DM between {\tt AGNdT8-9} and {\tt ViscHi-Lo}.The solid lines are for {\tt AGNdT8} and {\tt ViscHi}, and the dashed lines are for {\tt AGNdT9} and {\tt ViscLo}, respectively. The 2pCF is boosted for {\tt AGNdT8} and {\tt ViscHi}, therefore they have more clustering of matter than their counterparts. The solid black line is the 2pCF from the EAGLE reference model at $z=0$, while black dashed line represent the 2pCF at $z=0$ with the AGN feedback turned off.}
    \label{fig:fig4}
\end{figure}

\begin{figure}
    \centering
    \includegraphics[width=1\linewidth]{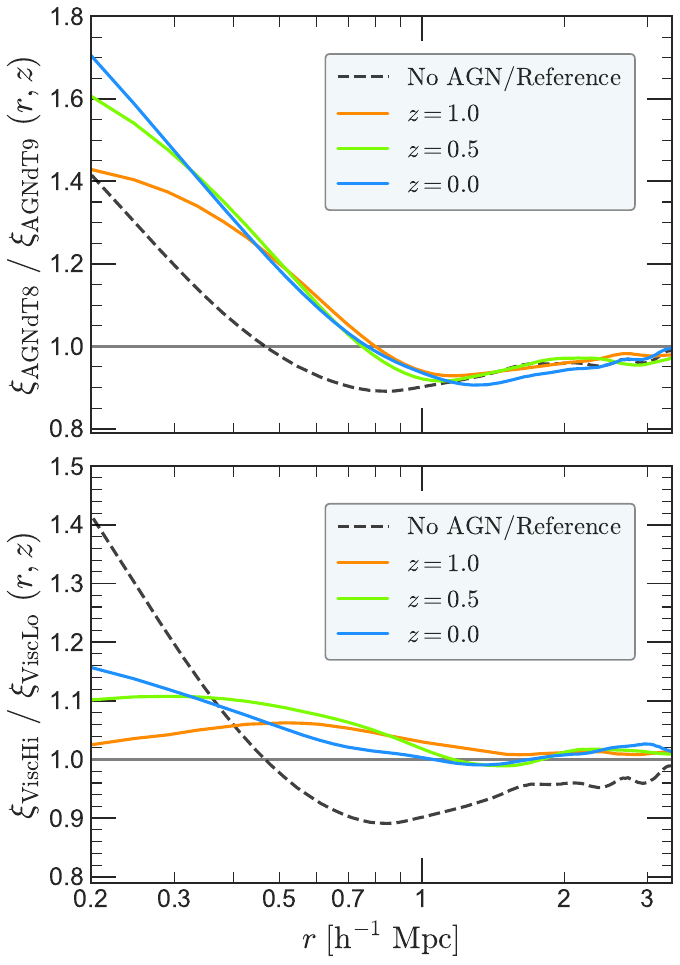}
    \caption{
    {\bf Top subplot}: {\tt AGNdT9/AGNdT8} 2pCF ratio shows $\sim50\%$ more suppression at small scales and $10\%$ boost at intermediate scales (at $z=0$) because of more efficient AGN feedback. {\bf Bottom subplot}: Even after early onset of AGN feedback  in {\tt ViscHi} 2pCF is boosted $\sim12-15\%$ in {\tt ViscHi} compared to {\tt ViscLo}. The solid black line represents the ratio of 2pCF, The comparison of the Reference model and that without AGN is shown for $z=0$ only.}
    \label{fig:fig5}
\end{figure}

Before we look into the quantitative differences between the different models, we first look at the qualitative differences. In Figure~\ref{fig:fig2}, we look at the temperature fields of the gaseous component for the different models at $z=0$ projected along the $z$-axis. The first row shows the gas temperature fields  for {\tt AGNdT8,9} model, while the bottom row shows the difference in the temperature fields for {\tt AGNdT8-9} and {\tt ViscHi-Lo} in a $10~\mathrm{Mpc~h^{-1}}$ region centred on the most massive halo in the box. Note that we only show the temperature difference field for the comparison of {\tt ViscHi/Lo} model to highlight their subtle differences more clearly.

From the first row, we notice that gas in {\tt AGNdT9} is hotter as well as more dispersed compared to the {\tt AGNdT8} model i.e when one looks at the bubbles in the same region, for e.g upper right corners, bubbles in {\tt AGNdT9} occupy larger area compared to {\tt AGNdT8}. Additionally, the black, empty regions are less prominent in {\tt AGNdT9} compared to {\tt AGNdT8} which implies hotter gas is dispersed over larger regions in {\tt AGNdT9} compared to {\tt AGNdT8}. Even in the zoomed region around the most massive halo, we see that gas in {\tt AGNdT9} is typically hotter. We observe that near the centre of the halo, gas in the {\tt AGNdT8} is hotter than {\tt AGNdT9} while, as we move away from the center, {\tt AGNdT9} is hotter than {\tt AGNdT8} as {\tt AGNdT9} has higher heating temperature than {\tt AGNdT8}, which leads to higher energy injection into the surrounding circumgalactic medium (CGM). Due to this higher energy injection, we expect gas heated by AGN to be transported further in {\tt AGNdT9} model. Hence, we see cooler gas in {\tt AGNdT9} compared to {\tt AGNdT8} around the centre of the halo, but as we move further, the temperature of the gas increases more rapidly in {\tt AGNdT9}. This also provides initial indication that gas in {\tt AGNdT9} is less clustered than {\tt AGNdT8}.

Next, we look at the zoomed region for {\tt ViscHi-Lo}. In {\tt ViscLo}, subgrid viscosity is lowered, which results in the delayed onset of AGN feedback. As a result more time is available for the gas to cool down in the CGM. Moreover, the energy injection by BHs in the {\tt ViscLo} model is lowered compared to the {\tt ViscHi} model. Therefore, we expect the matter in {\tt ViscLo} model to be more clustered, at least on the small scales compared to {\tt ViscHi}.

Now we look at a quantitative analysis of the gas distributions in the different models. We first look at the two-point statistics for different models at different epochs. The two-point correlation function (2pCF) and power spectrum ($P(k)$) are the most commonly used two-point statistics for quantifying the clustering of matter. The 2pCF and $P(k)$ is often used interchangeably, as they qualitatively convey the same information.

As the Universe evolves, the strength of clustering should increase, and this remains true for all the components of the Universe we study: gas, dark matter (DM), and total matter (gas + DM + stars + BHs). This is expected from these statistics, because, as the universe evolves, clustering on small scales increases as new stars form, and as gas accumulates inside galaxies (i.e. it is trapped inside dark matter potential wells). This is demonstrated in Figure~\ref{fig:figA3}, where we show the evolution of $P(k)$ for the DM and total matter component. 

In Figure~\ref{fig:fig3}, we show the ratio of power spectrum, $P(k)$, of DM with respect to gas for the {\tt AGNdT8} and {\tt ViscHi} models, at different redshifts, $z$. This demonstrates the effect of baryonic physics, particularly AGN feedback on the distribution of gas, i.e the suppression of $P(k)$ of gas relative to DM.
Interestingly, while the general trend is for ratio to increase (i.e increase in suppression of $P(k)$ of gas relative to DM) with decreasing redshift, this behaviour is not monotonic across all scales. In particular, we observe ``crossover'' points, i.e. where $P_{\rm DM}/P_{\rm gas}(k, z= z_i) < P_{\rm DM}/P_{\rm gas}(k, z = z_{i-1})$ for certain epochs. These crossover indicate where the clustering of gas has increased at some scales, relative to the general trend. This is a signature of AGN feedback, which redistributes matter from small scales to larger scales and thus increases the clustering of gas on larger scales. In Figure~\ref{fig:fig3}, we observe that there are crossover events at scales $k < 20~{h~\rm Mpc^{-1}}$. We observe the same with the {\tt AGNdT9} and {\tt ViscHi} models. It is notable that the crossover happens only for $z<2.0$, which further indicates the time at which the dominance of AGN feedback sets in (i.e. the black holes have grown large enough that they can significantly affect the matter distribution). This motivates us to look at the $P(k)$ with a finer time resolution between $z=1.5 - 0$.

In the two panels of Figure~\ref{fig:fig4}, we show the 2pCF of DM relative to gas, for {\tt AGNdT8-9} and for {\tt ViscHi-Lo} models at three epochs: $z=1, 0.5, 0$. We set the lower limit of $r$ to be $0.2~\mathrm{h^{-1}~Mpc}$, where the finite resolution of the simulations may be dominant. Further, we compare these to the EAGLE reference model \citep{Schaye_2015}, Table~\ref{tb:variation}, ran on the same box, with and without AGN Feedback. We make two significant observations from these plots:
\begin{itemize}
    \item The suppression of correlation function of gas, relative to DM, at small scales is more in {\tt AGNdT9} compared to {\tt AGNdT8}, i.e. matter in {\tt AGNdT8} is more clustered than {\tt AGNdT9} at small and intermediate scales, but at large scales $(r>1 h^{-1}~\mathrm{Mpc})$, the gas is slightly more clustered in the {\tt AGNdT9} model. This is true for all three epochs. We observe the same qualitative behaviour for $z>1$ as well.
    \item The suppression of correlation function of gas, relative to DM, at small scales is more in {\tt ViscLo} compared to {\tt ViscHi}, i.e. matter in {\tt ViscHi} is more clustered than {\tt ViscLo} at small scales and almost similarly clustered at the intermediate scales.
\end{itemize}

These observations may be understood as follows:\\
\textbf{First:} between {\tt AGNdT8-9}, $\Delta T_{\mathrm{AGN}}$ is varied, which controls the efficiency and energetics of AGN feedback (refer to the Section \ref{sec:AGN_feedback}). In {\tt AGNdT9}, AGN feedback is more efficient and energetic than the feedback events in {\tt AGNdT8}. Thus feedback events in {\tt AGNdT9} are better able to transport the gas from the center of the halos to the ICM, thus resulting in lower clustering at smaller scales, as feedback makes the halos less dense. The lower the radiative loss of gases within the halos, the greater is the efficiency of  feedback, thus resulting in less cooling, which quenches the star formation in galaxies and further reduces the 2pCF. The cumulative impact of these effects is to reduce the amplitude of the gas and all matter 2pCFs in the {\tt AGNdT9} compared to {\tt AGNdT8}.  \cite{Brun_2014} used the cosmo-OWLS simulation suite to conduct a systematic examination of the 
 properties of galaxy groups in response to variation of  $\Delta T_{\mathrm{AGN}}$. They too concluded that a higher heating temperature yields more efficient AGN feedback.

At intermediate scales,  we see that suppression in the 2pCF of gas relative to DM for {\tt AGNdT8} is less than {\tt AGNdT9} i.e there is a boost in the amplitude of 2pCF(gas) of {\tt AGNdT9}, resulting in a crossover of the ratios with {\tt AGNdT8}. This is an observational effect of AGN feedback, where the hot gas transported from the center of the galaxies and halos cools down at the intermediate scales ($\sim$few 100s of kpc), this increases the clustering at intermediate scales, and we observe an increase of amplitude of 2pCF of gas. BHs in {\tt AGNdT9} are more efficient than BHs in {\tt AGNdT8}, and are thus able to drive more material from the center of the galaxies to ICM; also, they can drive them further than BHs in {\tt AGNdT8}. This explains why the amplitude of 2pCF is larger in {\tt AGNdT9} than {\tt AGNdT8} at intermediate scales. In-fact we observe that the clustering of gas is more than the clustering of DM at intermediate scales for {\tt AGNdT9}

\textbf{Second:} between {\tt ViscLo} and {\tt ViscHi}, only the $C_{\mathrm{visc}}$ parameter is varied, which controls subgrid viscosity. Since AGN feedback suppresses the 2pCF at smaller scales, the suppression will be large in the case when the duration of the AGN feedback is longer. The {\tt ViscHi} model has higher kinetic viscosity (i.e lower $C_{\mathrm{Visc}}$), which results in early onset of AGN feedback; thus we expect {\tt ViscHi} to be less clustered than {\tt ViscLo}. Instead, the opposite trend is observed -- we return to this point later in this subsection.

In Figure~\ref{fig:fig5} we show the ratios of the 2pCF for the  model variations to have a better quantitative understanding of their differences. At $z=0$, there is almost $70\%$ stronger suppression in 2pCF in the {\tt AGNdT9} model compared to the {\tt AGNdT8} model at small scales, while at intermediate scales, there is boost of $10\%$ in the {\tt AGNdT9} model compared to {\tt AGNdT8} model. Despite the early on-set of AGN feedback in the {\tt ViscHi} model we observe a  $15\%$ boost in the 2pCF of {\tt ViscHi} model compared to {\tt ViscLo} model.

We also note from Figures~\ref{fig:fig4} and~\ref{fig:fig5} that the 2pCF in the Reference EAGLE model is bracketed by the {\tt AGNdT8-9} models, and also by the {\tt ViscHi-Lo} models, although the deviation is much more prominent in the case of {\tt AGNdT8-9} models. We also note that the 2pCF from {\tt ViscHi} model is very similar to the Reference EAGLE model, even after large change in the $C_\mathrm{Visc}$ parameter. This indicates that some parameters can have relatively little effect on the distribution of gas.
 
\begin{figure}
    \centering
    \includegraphics[width=1\linewidth]{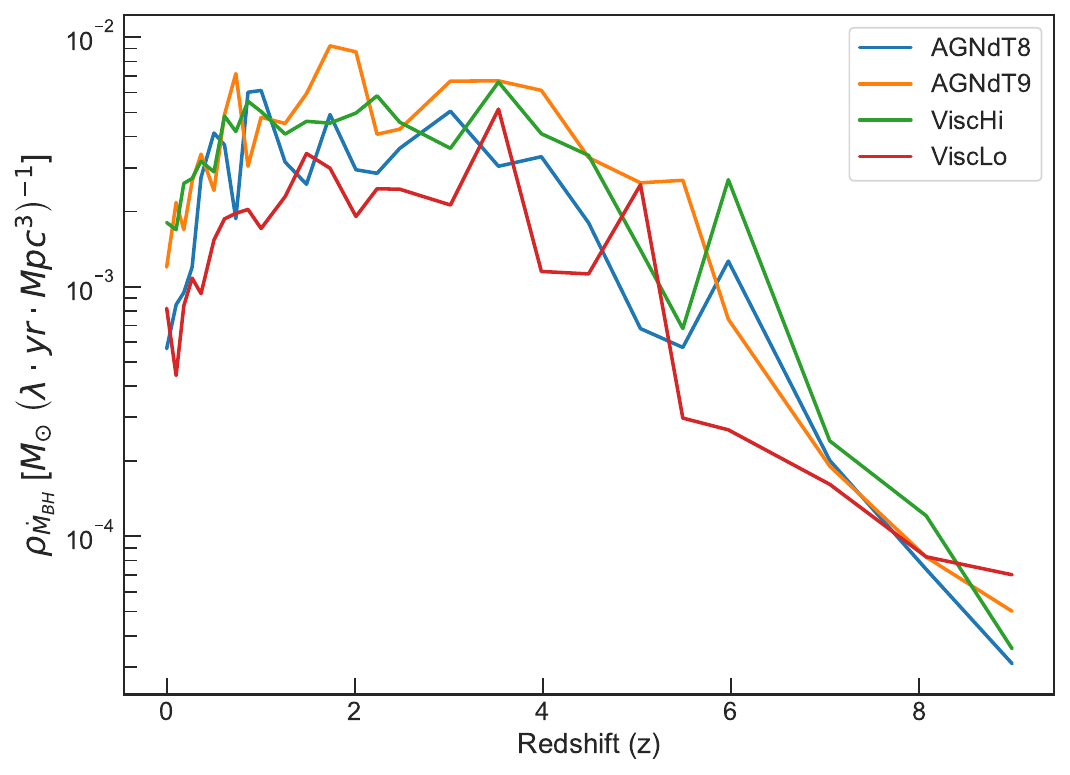}
    \caption{Black hole accretion rate density at different epochs. Higher viscosity ({\tt ViscHi}) leads to higher accretion rate, thus a higher energy injection into the surrounding than {\tt ViscLo}. BHs in {\tt AGNdT9} are more effective than BHs in {\tt AGNdT8}, thus they have higher accretion rate so that they can grow properly. Here $\lambda$ is a constant, which specifies the mass of accretion in solar masses.}
    \label{fig:fig6}
\end{figure}
\vspace{2ex}
\begin{figure*}
    \centering
    \includegraphics[width=1\linewidth]{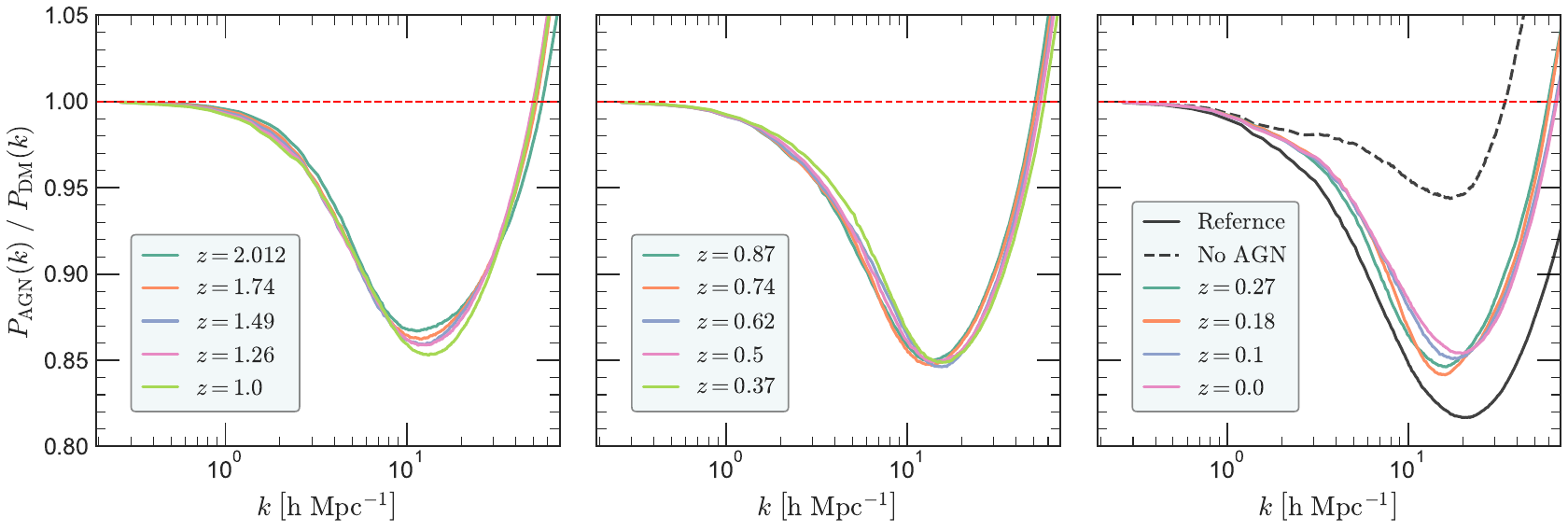}
    \caption{Ratio of the total matter and the corresponding DM power spectrum at different epochs for {\tt AGNdT8} model. The solid black curve shows the same ratio for the EAGLE Reference model at $z=0$ while the dashed black line shows the same ratio with AGN feedback turned off at $z=0$.}
    \label{fig:fig7}
\end{figure*}

\begin{figure*}
    \centering
    \includegraphics[width=1\linewidth]{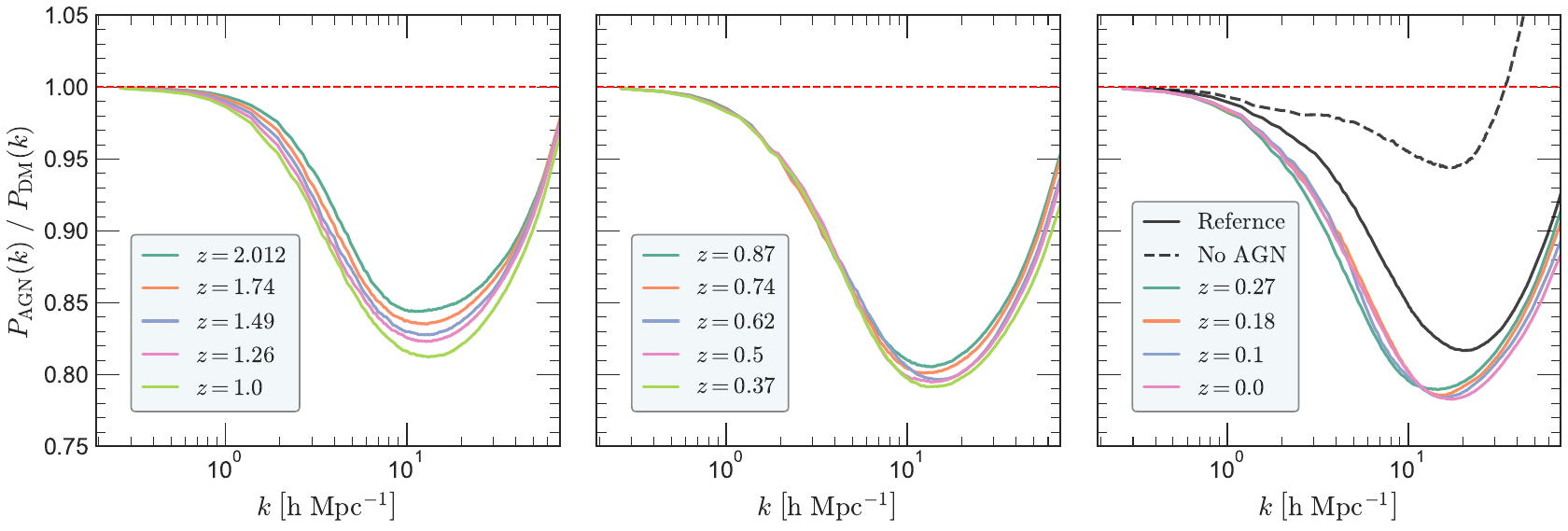}
    \caption{Ratio of the total matter and the corresponding DM power spectrum at different epochs for {\tt AGNdT9} model. The solid black curve shows the same ratio for the EAGLE Reference model at $z=0$ while the dashed black line shows the same ratio with AGN feedback turned off at $z=0$.}
    \label{fig:fig8}
\end{figure*}

\begin{figure*}
    \centering
    \includegraphics[width=1\linewidth]{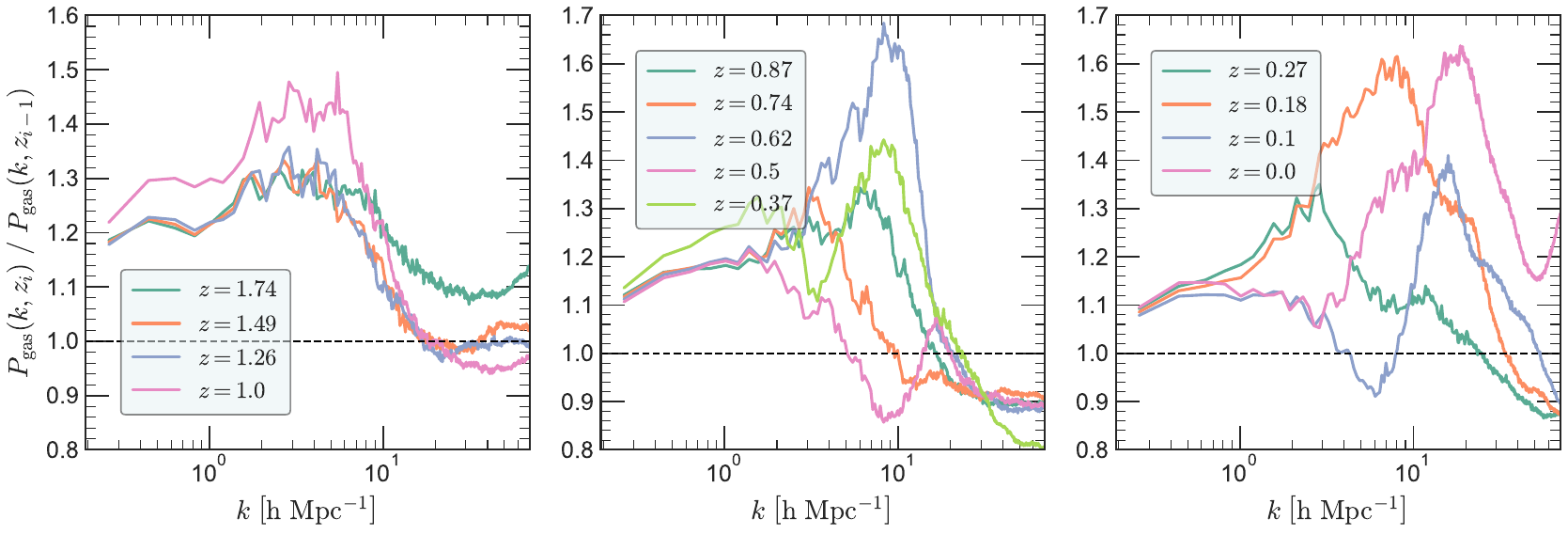}
    \caption{Ratio of power spectrum of gas between two consecutive epochs for {\tt AGNdT8} model. The panels are arranged in decreasing redshift if we go from left to right. The ratio significantly falls below 1 at $z=1$ (first panel from left) which marks the onset of dominance of AGN feedback. }
    \label{fig:fig9}
\end{figure*}

We noted previously that the {\tt ViscHi} model exhibits a somewhat more clustered gas distribution despite the physics of the model which is expected to lead to the earlier onset of AGN feedback. To understand the reason behind this, we examine the activity of black holes in these models  in detail. To this end, we sum the black hole accretion rate at a given epoch and divide it by the volume of the box. In other words, this corresponds to the \textit{total} BH accretion rate density measured in the simulation.

The results are shown in Figure~\ref{fig:fig6}, where we can see that BH accretion rate density is higher in {\tt AGNdT9} and {\tt ViscHi} models compared to their counterparts, {\tt AGNdT8} and {\tt ViscLo}. BHs in {\tt AGNdT9} are more active than BHs in the {\tt AGNdT8} model; a higher accretion rate density for {\tt AGNdT9} therefore agrees with our earlier observations. A higher accretion rate density results in more redistribution of gas, and hence suppression in the 2pCF.  The {\tt ViscLo} model has larger $C_{\mathrm{Visc}}$ parameter which results in lower accretion rate of BHs (Equation \ref{eq:accr}) compared to BHs in the {\tt ViscHi} model. \cite{Crain_2015} found that having a higher subgrid viscosity i.e lower value of $C_{\mathrm{Visc}}$ parameter leads to higher energy injection rate when the accretion is in viscosity-limited regime. A higher energy injection rate leads to stronger AGN feedback, and hence more redistribution of matter. This suggests that the implication of the change of accretion rate through $C_{\mathrm{Visc}}$ is far more complex, and the observed correlation function is due to the complex interplay of redistribution of gas through AGN feedback and the interaction of the gas with its surrounding. In this sense, it is harder to develop a direct mapping between the expected effect on the clustering of gas based on changed in subgrid viscosity than it is, for example, with changes in the AGN heating temperature.

From the above discussion, it is clear that $\Delta T_{\mathrm{AGN}}$, which controls the heating of particles around AGN is more influential than viscosity of gas in reshaping the matter distribution on intermediate and large scales. Thus for the rest of the analysis, we will consider just the {\tt AGNdT8} and {\tt AGNdT9} models.

\subsection{Time constraints on the dominance of AGN Feedback}\label{subsec:time}
AGN feedback redistributes gas. The DM component remains largely unaffected by changing the AGN feedback parameters. From our previous analysis in Section \ref{subsec:models} we have seen that the dominance of AGN feedback manifests as crossover points in the power spectrum on small scales. Thus, to put further constraints on the effects of AGN feedback, we first look at the ratio of $P(k)$ between the total matter and DM at different epochs using finer snapshot spacing between consecutive epochs. This enables us to understand the temporal and spatial onset of AGN feedback in the EAGLE model.

\begin{figure*}
    \centering
    \includegraphics[width=1\linewidth]{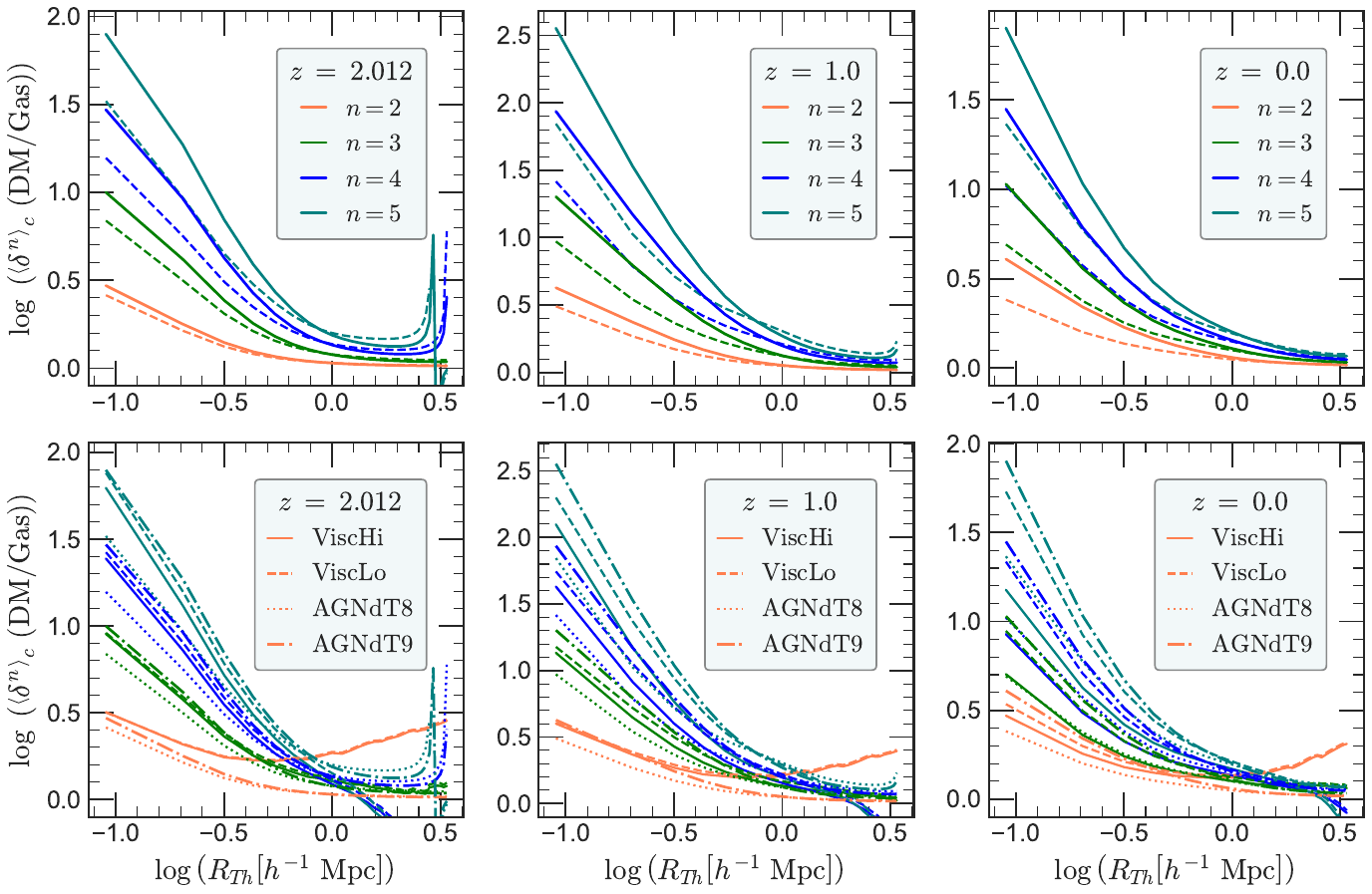}
    \caption{Cumulants of the DM density field relative to gas at different epochs for different smoothing scales. The color of the lines indicates the curve while the line styles indicate the different models in our study. The top panels show the suppression in the cumulants for {\tt AGNdT9} (solid curves) and {\tt AGNdT8} (dashed curves) models. In the bottom panels for completeness, we plot all the models.
    }
    \label{fig:fig10}
\end{figure*}

From Figures~\ref{fig:fig7} and~\ref{fig:fig8}, we observe the same features that we observed in Figure~\ref{fig:fig4}, i.e., there is stronger suppression of $P(k)$ in {\tt AGNdT9} compared to {\tt AGNdT8}, which is explained in the discussion following Figure~\ref{fig:fig4}. Additionally, we also observe that suppression between each consecutive epoch is more in the case of {\tt AGNdT9} than {\tt AGNdT8} model, which is due to the higher efficiency of AGN feedback in {\tt AGNdT9}. We also note that the baryonic effects are visible on scales as large as $0.8~\mathrm{h~Mpc^{-1}}$ even at $z\approx2$. In Figures~\ref{fig:fig7} and~\ref{fig:fig8} we also plot the same ratios for the EAGLE Reference model (solid black curve) and a model variation with AGN Feedback turned off (dashed black curve), both at $z=0$. We again note that the EAGLE Reference model is bracketed by the {\tt AGNdT8-9} models. Moreover, even turning off the AGN feedback results in a  change in the $P(k)$ by $\approx 5\%$ at the intermediate scales ($\approx10~ h~\mathrm{Mpc^{-1}}$). Interestingly, the suppression in $P(k)$ for all models is most pronounced on scales $k\approx15-20\,h$ Mpc$^{-1}$, with the maximal suppression (relative the dark matter-only case) being of the order of $20-25\%$. These observations further demonstrate the need to better understand AGN feedback to accurately model the matter distribution at scales, $ k > 1~h~\mathrm{Mpc^{-1}}$ for precision cosmology applications.

Finally, we look at the crossover events to determine when AGN feedback started to dominate. We plot the ratios of the power spectrum of gaseous component between two consecutive epochs and check when the ratio significantly falls below 1. The result is presented in Figure~\ref{fig:fig9} for {\tt AGNdT8}. We can see that the ratio significantly falls below one at $z=1$. As we go to lower redshifts, we see that the decrease of the ratio becomes sharper, and we start to notice the fall at intermediate scales as well. The suppression at intermediate scales appears first at $z=0.74$ and is prominent by $z=0.5$. Thereafter, the effects of AGN feedback start to equilibriate, and for $z<0.5$, we notice the magnitude of slope decreases, and for some epochs, we do not see the ratio falling below one. These observations suggest that, in the EAGLE model, AGN feedback starts to dominate at $z\approx 1 - 0.7$, and after that, effects start to equilibriate with the surrounding. For the {\tt AGNdT9} model, we observe that AGN feedback starts to dominate somewhat earlier, $z=1.49-1$, due to increased efficiency of BHs. \cite{2017McAlpine} found similar results by comparing the star formation rate (SFR) and black hole accretion rate in EAGLE Reference Model. Since AGN feedback quenches the star formation in galaxies, SFR can be used as an indirect tracer for feedback from BHs. They found a rapid decline in the SFR in halos with $M_{200}\approx 10^{12}M_\odot$ which has a median redshift, $z=1.9$, which is very close to our inference from the evolving $P(k)$.

\subsection{Cumulants of the matter distribution}
As described in Section~\ref{subsec:cumulants}, we use cumulants of the matter distribution to probe the non-linear evolution of the matter density field. We compute the cumulants at different epochs. We first compute the density field for the DM and gas components of the matter. Then we smooth the density field with a Gaussian kernel and compute the cumulants. We repeat the step for different values of the smoothing scale, $R_{\rm Th}$, and at different redshifts. For our analysis, we use\break $\dfrac{2}{752}\mathrm{BoxSize} < R_{\rm Th} < 0.1 \mathrm{BoxSize} $ where $\mathrm{BoxSize}$ is the size of the simulation box and is equal to $33.885~h^{-1}~\mathrm{Mpc}$. The lower bound is set using the Nyquist frequency, while the upper bound is set empirically. In Figure~\ref{fig:fig10} we plot the ratio of cumulants of the DM w.r.t. gas at different epochs, as a function of $R_{\rm Th}$, for {\tt AGNdT8} and {\tt AGNdT9}.

Cumulants are an essential tool to detect deviations from Gaussianity. A Gaussian field is characterised by only the first two moments or cumulants, and higher-order cumulants are zero; non-zero higher-order cumulants therefore indicate non-Gaussian behavior. Larger deviations from Gaussian behaviour is indicated by larger values of high order $\Fb{n>2}$ cumulants. When this non-Gaussian field is smoothed using a Gaussian kernel, the non-Gaussian characteristic gradually dampens as we increase the size of smoothing kernel, $R_{\rm Th}$. 
Indeed, it is expected that for large enough values of $R_{\rm Th}$, all the higher-order cumulants should converge to zero, and only the variance (i.e., the second-order cumulant) should survive. 

We have already established that as the universe evolves and structure formation takes place, the strength of clustering increases for both DM and gas. Thus, as $z$ decreases, the irregularities grow stronger and stronger; therefore we expect the amplitude of the cumulants to increase for all orders. However, baryonic feedback re-distributes gas which decreases the clustering strength i.e. same as lowering of the density contrast of the gas relative to DM -- which implies a lower amplitude for the cumulants of gas -- and hence similar to 2pCF or $P(k)$, we will observe a suppression in cumulants of gas relative to DM for all orders.

In the top panel Figure \ref{fig:fig10},  we show the cumulants of gas for {\tt AGNdT8} (dashed curve) and {\tt AGNdT9} (solid curve) relative to the DM for different redshifts. We we observe that ratio is enhanced for {\tt AGNdT9} compared to {\tt AGNdT8} for all orders and across redshifts. This is because BHs in the {\tt AGNdT9} model are more efficient than BHs in the {\tt AGNdT8} model, hence in the {\tt AGNdT9} model,  gas particles are distributed over larger distances in the simulation box, thereby making the gas distribution is more uniform in {\tt AGNdT9} compared to the {\tt AGNdT8} model, resulting in more suppression of amplitude of cumulants compared to DM in the {\tt AGNdT9} model. We also observe that while at $z=2.012$, there is a significant difference in the ratios in the higher order cumulants (i.e. $n>2$) between {\tt AGNdT8/9}, the difference for second cumulant (i.e. $n=2$) is very small. The difference in the second order cumulants only grows for $z \leq 1$, which can be thought of as a constraint on the time scale at which AGN feedback starts to alter the underlying distribution of gas field relative the DM field - this is agreement with the constraints obtained in Sec \ref{subsec:time}. We also observe that as we increase the size of the smoothing kernel, the ratio starts converging to 1, i.e the DMO and hydrodynamical runs have similar amplitudes for the cumulants.

In the bottom panel of Figure \ref{fig:fig10}, we show the ratio of cumulants (DM/gas) for all the models in our study (represented by the different line styles, while the color indicates the order of cumulants which is same as the top panels). First of all we observe that for the  {\tt ViscHi/Lo} models all the higher order cumulants $(n>2)$ demonstrate behaviour similar to the {\tt AGNdT8/9} case. However, the second order cumulants $(n=2)$ demonstrate a different behaviour -- particularly, with increasing smoothing radius, the second order cumulants for gas drops rapidly, which correspondingly increases the ratio. This suggests that changes to the viscosity parameter have a more complex impact on the variance of the gas density even on large scales. This maybe an interesting thread for further examination, but this is beyond the scope of this work. We also observe that {\tt AGNdT9}, shows the highest suppression of cumulants of gas relative to DM, and the {\tt AGNdT8/9} models shows larger difference in the suppression compared to {\tt ViscHi/Lo} models. This further motivates us to think that AGN heating temperature parameter $(\Delta T_{\rm AGN})$ has significant impact on the overall gas distribution.

\subsection{The three-point correlation function}
\label{sec:3pCF}
\begin{figure}
    \centering
    \includegraphics[width=0.95\linewidth]{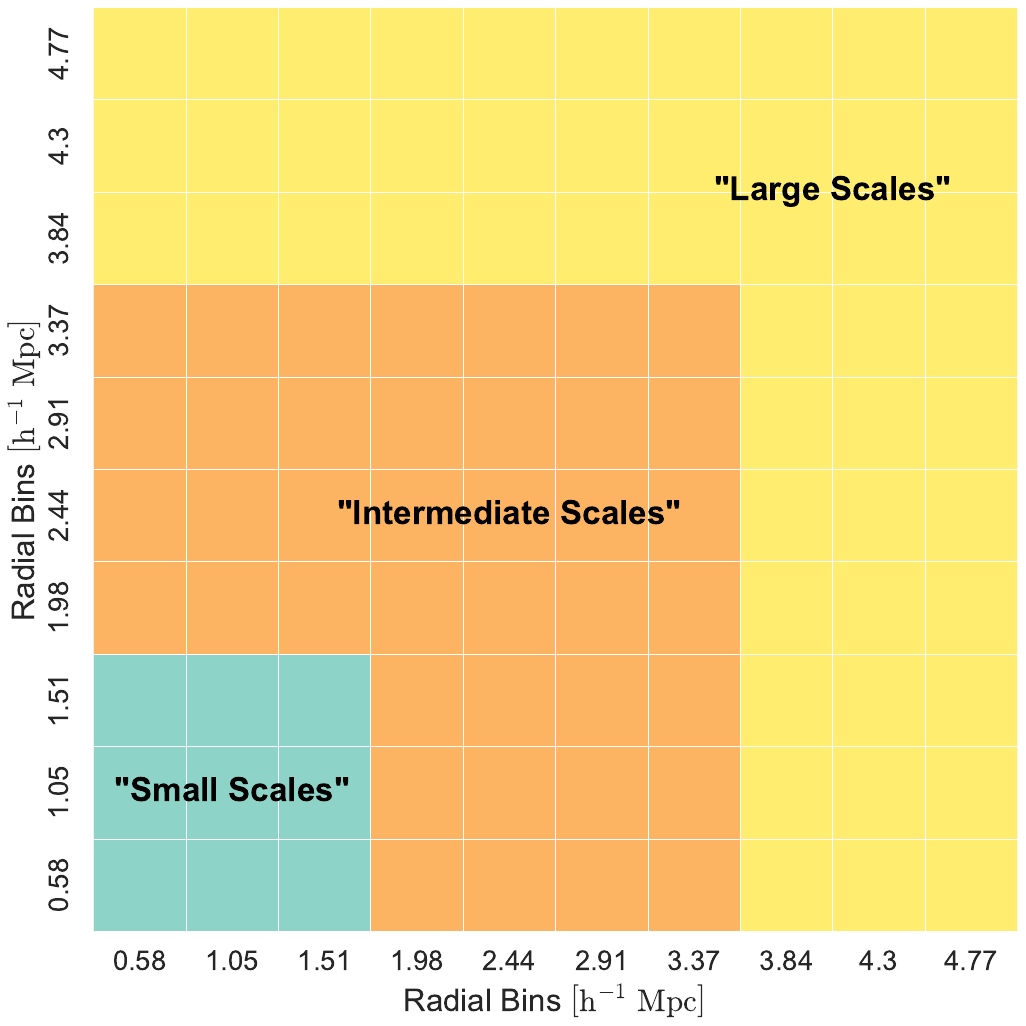}
    \caption{An illustrations depicting different scales in the 3pCF matrix.}
    \label{fig:fig11}
\end{figure}

\begin{figure*}
    \centering
    \includegraphics[width=1\textwidth]{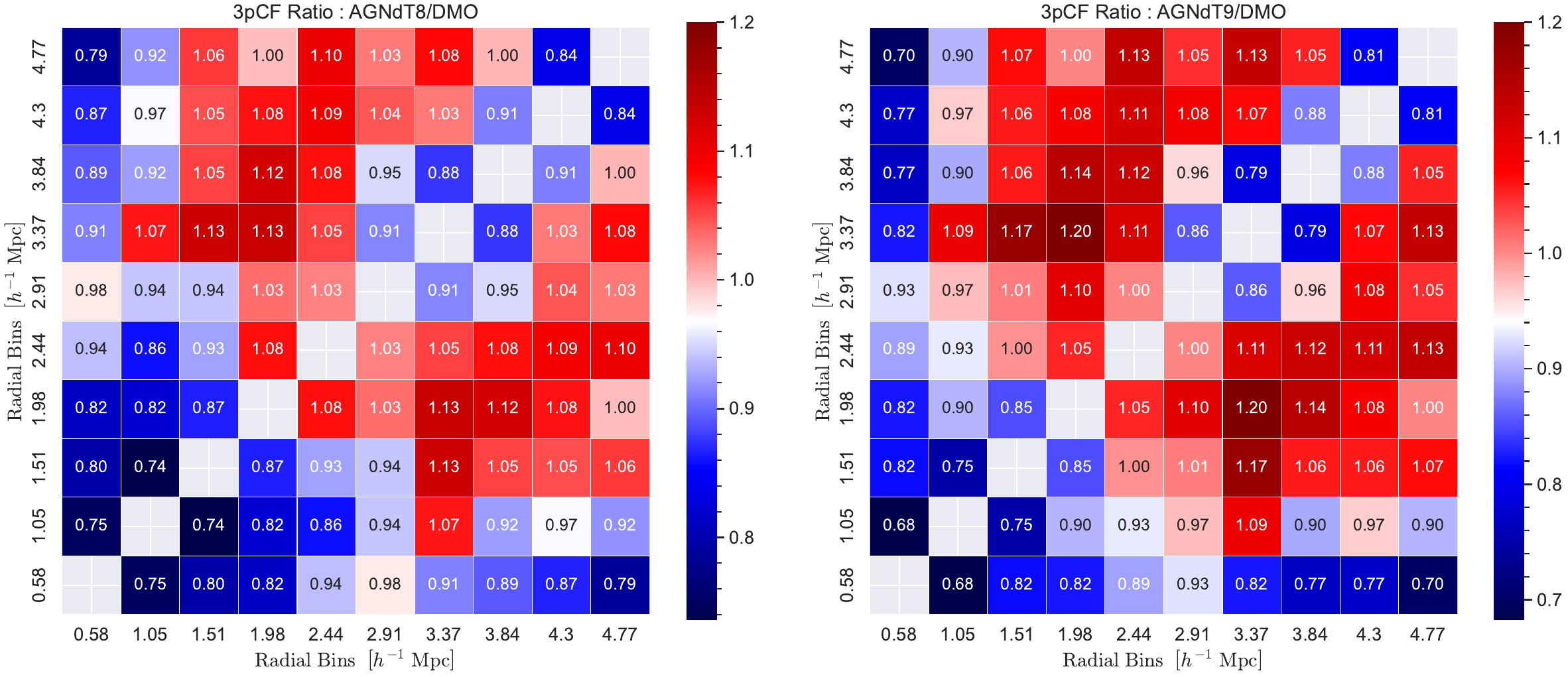}
    \caption{3pCF of gas in {\tt AGNdT8/9} relative to DM. AGN feedback redistributes the gas from the central part of the halos to the CGM, which reduces the clustering on small scales. Thus we see a suppression of 3pCF on small scales and boost at intermediate and large scales. The suppression and boost is more for the {\tt AGNdT9} model due to more efficient AGN feedback. 
    }
    \label{fig:fig12}
\end{figure*}

\begin{figure}
    \centering
    \includegraphics[width=1\linewidth]{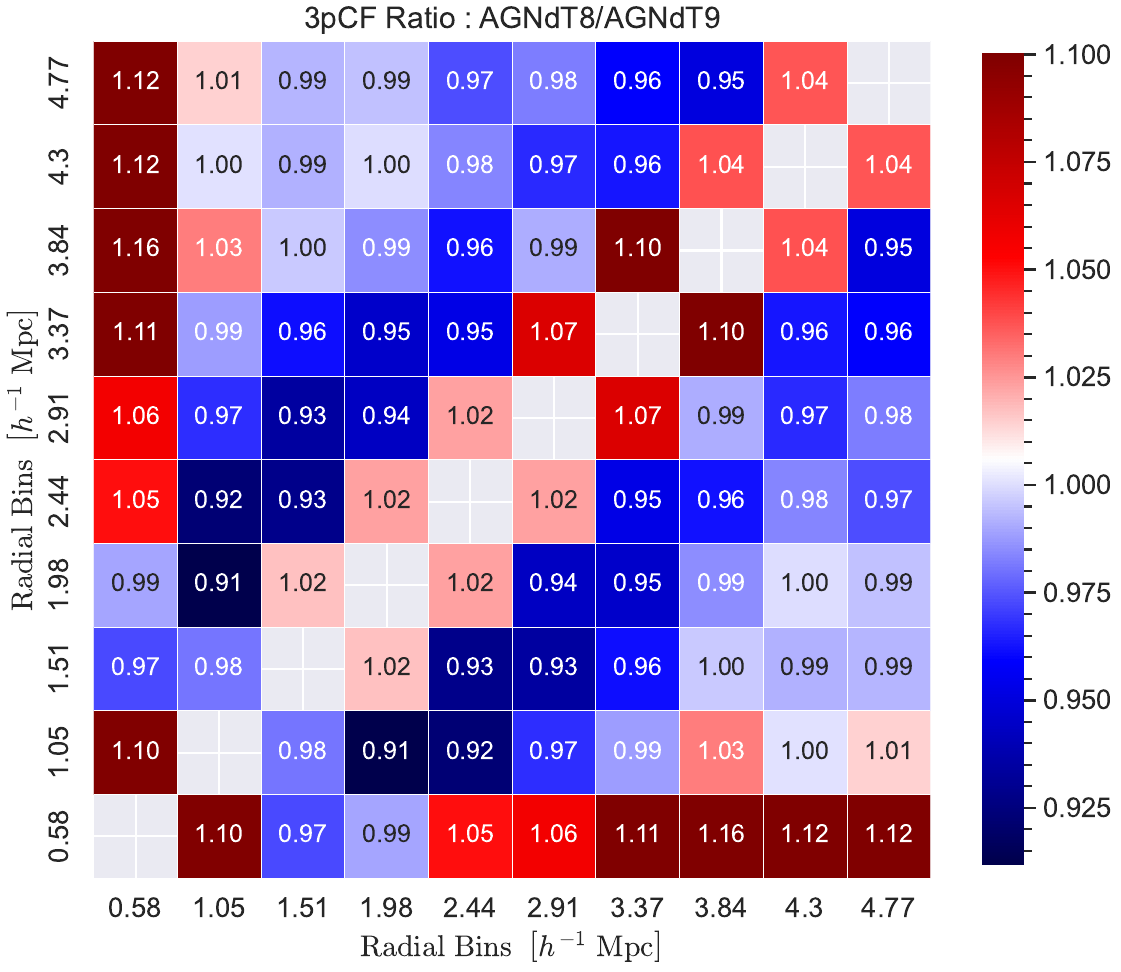}
    \caption{Ratio of 3pCF between {\tt AGNdT8, AGNdT9} model.}
    \label{fig:fig13}
\end{figure}

Compared to the 2pCF, the three-point correlation function (3pCF) is relatively understudied as a tool for quantifying the matter distribution in hydrodynamic simulations. This is largely due to the increased computational cost ($\approx\mathcal{O}(n^3)$ given $n$ tracers). However, the 3pCF may contain more information on the distribution of matter than what is simply contained in the 2pCF. This section discusses our result from the 3pCF analysis of {\tt AGNdT8}, {\tt AGNdT9}. 

Our observations from Figure~\ref{fig:fig10} already give us an initial impression that the non-Gaussian distribution of the gas changes if we vary the AGN feedback, as we observe that the amplitude of the higher order cumulants ($n>2$) are significantly different at small smoothing scales between {\tt AGNdT8} and {\tt AGNdT9}. However, it does not tell us anything about the strength of the changes because cumulants do not consider the correlation between points. This motivates the use of the 3pCF. In order to compute this from our simulation data, we first take 20 random uniform samples from the entire dataset containing $0.1\%$ of the total gas particles; this is done to reduce computational cost. This step does not affect our conclusions; we have checked explicitly that our results are converged with respect the size of our sub-sample (Appendix~\ref{sec:3pCF-Sampling}). We then compute the 3pCF on each of these datasets, and then obtain the final 3pCF matrix by taking the average of the 3pCFs estimated from the subsets. We consider triangles with sides up to 5 Mpc only, and we bin the triangles into ten bins per side. We then estimate the error in the 3pCF value for a given radial bin by taking the standard deviation between the 3pCF for the same radial bin from the samples. 

In Figure~\ref{fig:fig11}, we present an illustration of the 3pCF matrix, highlighting the different scales where the information content is stored. The $x$ and $y$ ticks are the average of the upper and lower bound of the radial bin, so in given bin $(r_1,r_2)$, we consider triangles whose sides length $(r_1 \pm \Delta r, r_2\pm\Delta r)$ Mpc, where $\Delta r=0.25$ Mpc. We loosely define ``small scales'' as the regime where galaxies may dominate i.e triangles with side lengths up to $1.7~\mathrm{Mpc}$. ``Intermediate scales'' denote regions of a few virial radii of the most massive haloes in the simulation i.e., galaxy clusters with side lengths upto $3.6~\mathrm{Mpc}$. Finally, ``large scales'' are defined as anything beyond this scale. We caution that these demarcations are determined qualitatively, and there is no universally accepted limit on the choice of different scales.

In Figure \ref{fig:fig12} we present the ratio of 3pCF for {\tt AGNdT8/9} model and compare it with the 3pCF from the DMO runs, whereas in Figure \ref{fig:fig13} we show the ratio of 3pCF between {\tt AGNdT8/9}. To first order, we observe that on small scales, the amplitude of the 3pCF of the gas from the  hydrodynamical runs is lower compared to the 3pCF from the DMO run. On the other hand, at the intermediate and the large scales, the 3pCF is boosted compared to the DMO case -- this is one more signature of AGN feedback, where gas is redistributed from the small scales to the large scales. This redistribution of gas lowers the DDD count in the hydrodynamical runs relative to the DMO run at small scales, while it is  boosted at the intermediate and the large scales. We also observe that this suppression of the 3pCF on small scales is greater for {\tt AGNdT9} compared to {\tt AGNdT8}, while the boosting at intermediate and large scales is more in {\tt AGNdT9}

From Figure~\ref{fig:fig13}, which compares the 3pCF of {\tt AGNdT8} and {\tt AGNdT9}, we see that the 3pCF of the two models are similar in magnitude. The 3pCF of {\tt AGNdT9} is slightly higher at the intermediate scales. We argued in Section~\ref{subsec:models} that BHs in {\tt AGNdT9} are more efficient than BHs in {\tt AGNdT8}, i.e they have less radiative loss, which results in more effective gas transport from the centre of the halos to intermediate scales (i.e., CGM), which results in higher DDD count at intermediate scales in the {\tt AGNdT9} model. The same argument holds for large scales as well; however, on large scales, this change is very negligible, so the ratios for most of the triangle configuration remain $\approx$ 1.
Additionally, if we look at the edge case (one side of the triangle is fixed at the smallest value allowed in calculation i.e. the bottom most row in the 3pCF matrix), it qualitatively reproduces the ratio curve similar to Figure \ref{fig:fig5} for the {\tt AGNdT8/9} case at $z=0$.
It is reassuring to see that the structure of the 3pCF matrix (at least qualitatively) provides complementary information to the 2pCF, while agreeing with the general trends observed previously.

For completeness, we also checked the 3pCF from the {\tt ViscHi-Lo} model and from these model also, we reach the same conclusion. However, the difference in the magnitude between the two models is very less ($\approx 2\%$ on average in a given radial bin) compared {\tt AGNdT8-9}, hence we opt not to show these results here.

\section{Discussion}
\label{sec:discussion}
We found that the effects of varying the kinematic viscosity have very tiny effect on the 2pCF or $P(k)$. Additionally, directly mapping the effects of the viscosity parameter with the resulting impact on the gas through feedback is rather difficult. Using EAGLE, we can only study the effects upto $3.3~\mathrm{h^{-1}~Mpc} \approx 5~\rm{Mpc}$. It might be the case that effects of viscosity manifest on scales larger than this. Therefore to understand the effect of viscosity we need simulations with larger box size and with higher resolution, with much more drastic variations in the viscosity. Moreover, to study the large-scale effects of changing these parameters, we need a much bigger simulation box size, at least of the orders of 100 $\mathrm{Mpc}$,  that would enable us to measure the effect on the 2pCF on the scale of 10s of Mpc. Indeed, the subgrid treatment of gas viscosity used in this work may be too crude, and fails to capture complex kinematics of gas flows around black holes.

The same caveat holds for the cumulants as well -- in principle at larger smoothing scales, the higher order cumulants should converge which we also observe in Figure~\ref{fig:fig10}. However it may be the case that cumulants measured in simulations with different AGN feedback prescriptions converge on scales larger than those probed in the simulations used here. That being said, however, the {\it qualitative} comparison for {\tt AGNdT8} and {\tt AGNdT9} remain valid. \cite{borrow2022impact} showed that globally averaged properties in a full cosmological volume differ between clone simulations, but the deviation diminishes for box size > 25 Mpc; this is because of stochasticity in the subgrid models underlying such simulations. For more robust constraints on measurements involving two-point statistics, we need multiple runs with the same configuration (i.e., cosmological and galaxy formation models) to enable better statistical fidelity in our estimates of the cumulants.

It may also be the case that the size and resolution of the simulations we use are not enough to take full advantage of the 3pCF. In principle, the three-point function should contain more information about changes to the large-scale matter distribution due to the effects of feedback than just $P(k)$ or the 2pCF, but this is not immediately apparent based on our observations in Section~\ref{sec:3pCF}. The 3pCF estimator is more likely to be affected by our limited sample size than the 2pCF and it will therefore be interesting to revisit this exploration with larger volume hydrodynamical simulations, particularly those that also vary subgrid parameters relating to feedback more widely than the set considered in this work.

Nevertheless, in this work we have noted how changing the strength of AGN feedback is able to introduce change in the two-point statistics and higher-order cumulants by several tens of percentage even at scales $k\approx 1~\mathrm{h~Mpc^{-1}}$. Given that upcoming surveys are hoping to constrain cosmological parameters using observables measured on scales $0.1 \leq k \leq 10 \mathrm{h~Mpc^{-1}}$ to 1\% or better, this further strengthens the case for why we need a better understanding of the baryonic physics and feedback effects.

\section{Conclusions}
\label{sec:conclusion}
This work investigates the effects of AGN feedback on the large-scale matter distribution in cosmological, hydrodynamical simulations focussing on the redistribution of gas in and around dark matter haloes. In particular, we use variations in the subgrid physics model of the EAGLE simulations \citep{Schaye_2015, Crain_2015} to study the impact of varying the BH feedback model on two- and three-point statistics used to the characterise the matter distribution. Our main findings are summarised below:
\begin{enumerate}
    \item  We find that the efficiency of AGN feedback (controlled by the parameter $\Delta T_{\tt AGN}$, which determines how much the surrounding particles are heated) is crucial in shaping the gas distribution on large scales (Figure~\ref{fig:fig2}). More efficient AGN feedback results in a stronger suppression of the two-point correlation function (2pCF) and the  power spectrum, $P(k)$, at small scales ($r < 1$ $h^{-1}~\mathrm{Mpc}$) and enhances it at intermediate scales (1 $h^{-1}~\mathrm{Mpc}$ $< r <$ 10 $h^{-1}~\mathrm{Mpc}$, Figures~\ref{fig:fig3}~\&~\ref{fig:fig4}). This is because, more efficient AGN feedback leads to more gas transport from the centres of galaxies to the circumgalactic medium (CGM).
    \item Increasing the viscosity of gas while keeping the efficiency of AGN feedback fixed results in earlier onset of AGN feedback. Thus, one would naively expect to see the redistribution of gas from the centre of galaxies to the CGM for the model with higher viscosity and thus a reduced clustering at small scales. However, we instead observe the reverse effect -- that the higher viscosity model is more clustered than the low viscosity model (Figure~\ref{fig:fig4}). This warrants further investigation, potentially involving larger simulations than the ones we have used in this work.
    \item By considering the ratio of $P(k)$ for gas between two consecutive epochs, we are able to narrow down the redshift range during which effects of AGN feedback were most dominant as $z=1-0.74$ (Figure ~\ref{fig:fig9}). However, this time scale also varies depending on the efficiency of the BHs -- more efficient BHs i.e. in {\tt AGNdT9} results in the effects showing up earlier at $z=1.49-1$ (Figure~\ref{fig:figA4}).
    \item Varying AGN feedback does not change the cumulants of dark matter distribution because AGN feedback mainly affects the gas, and the any net effect on the DM is negligible. More efficient AGN feedback results in a less clumpy gas distribution, resulting in a reduction in the magnitude of the cumulants compared to the model with less efficient AGN feedback (Figure~\ref{fig:fig10}).
    \item The 3pCF of the gas distribution (Figure~\ref{fig:fig12}) also qualitatively exhibits similar behavior to the 2pCF. Due to re-distribution of gas from baryonic processes, the 3pCF is suppressed at smaller scales ($r<1.7 ~h^{-1}~{\rm Mpc}$) for the gas w.r.t to DMO runs, while it is boosted at the intermediate and large scales ($r > 1.7 ~h^{-1}~{\rm Mpc}$).
    \item For the physics variations used in this study, the 3pCF shows only minor changes for variations of the same type i.e. between {\tt AGNdT8-9} and {\tt ViscHi-Lo}. The only noticeable changes are  on intermediate scales, as more gas is transported from the centre of the haloes to the intermediate scales when the AGN feedback is more efficient. This boosts the 3pCF at intermediate scales. These observations are consistent with what we concluded with the 2pCF. The present simulations may, however, be too limited in the size and resolution, to extract the full information content in the 3pCF. 
\end{enumerate}
This paper adds to the growing body of work demonstrating the importance of considering the effects of galaxy formation and feedback on the large-scale matter distribution, particularly given the ambitions of precision cosmology. We have shown how relatively small variations in parameters that are, in general, poorly constrained can leave imprints on the matter distribution from anywhere between 5-25\%, depending on the model, redshift, and the scales of interest. Our work also shows some of the limitations of finite box size and the scope of model variations we have considered; in particular, it would be illuminating to consider the use of the 3pCF in characterising the gas distribution in larger simulations that also incorporate a more wide range of feedback mechanisms. New generations of hydrodynamical simulations like the FLAMINGO project \citep{schaye2023flamingo} provide the perfect opportunity to pursue these scientific questions.

\section*{Acknowledgements}
We thank the anonymous referee for providing suggestions that have improved the quality of our work. SB is supported by the UK Research and Innovation (UKRI) Future Leaders Fellowship [grant number MR/V023381/1]. This work used the DiRAC@Durham facility managed by the Institute for Computational Cosmology on behalf of the STFC DiRAC HPC Facility (www.dirac.ac.uk). The equipment was funded by BEIS capital funding via STFC capital grants ST/K00042X/1, ST/P002293/1 and ST/R002371/1, Durham University and STFC operations grant ST/R000832/1. DiRAC is part of the National e-Infrastructure.

\section*{Data Availability}
The data supporting the plots within this article are available on request to the corresponding author and can also be access from the following repository: \href{https://github.com/sparxastronomy/AGN_feedback/}{github.com/sparxastronomy/AGN\_feedback/}.


\bibliographystyle{mnras}
\bibliography{references} 


\appendix
\counterwithin{figure}{section}

\section{Two point Estimates}
\begin{figure}
    \centering
    \includegraphics[width=1\linewidth]{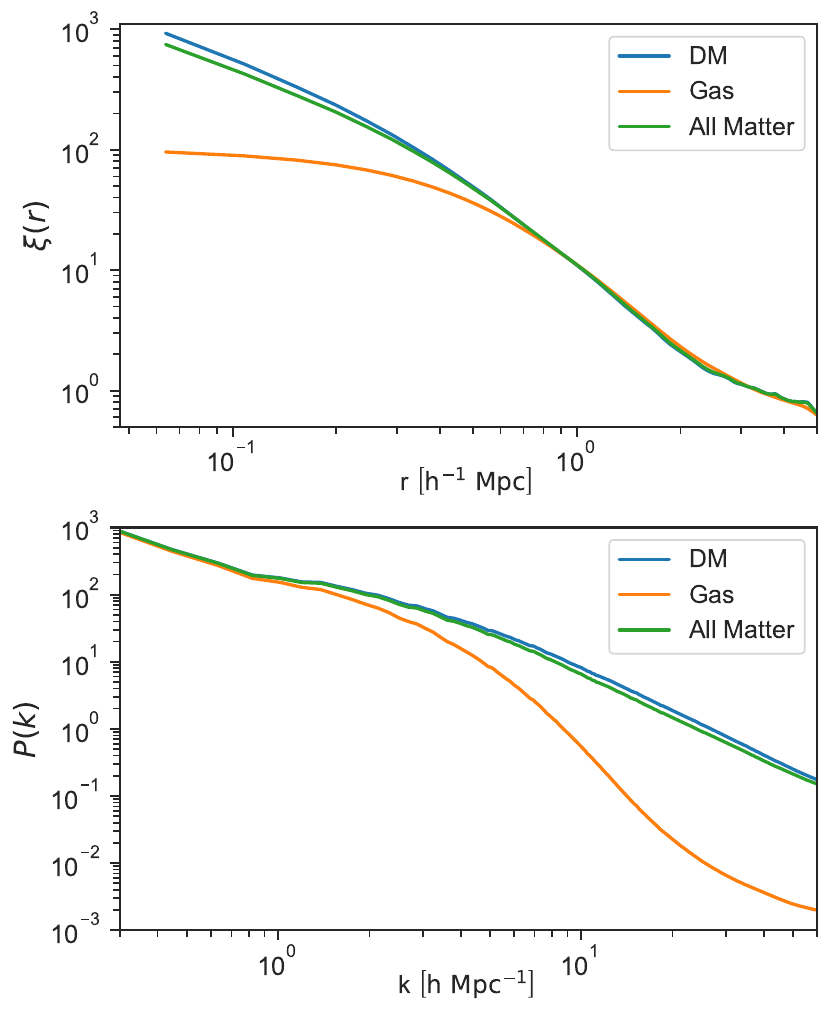}
    \caption{2pCF and $P(k)$ for {\tt AGNdT9} model. This figure compares the 2-point estimate for different components of the matter. The DM component remains invariant in all model variations. Only the gas and the total matter (DM + gas + stars + BHs) components get affected. We are interested in the change of gas distribution as that is the component that is most significantly affected by variations in the AGN feedback model.}
    \label{fig:figA1}
\end{figure}

In this section, we take a closer look at the 2-point estimates. Figure~\ref{fig:figA1} compares the 2pCF and $P(k)$ for different components of the matter distribution in the {\tt AGNdT9} model. We observe similar behavior in all the other models. The gaseous components of the matter distribution is affected the most due to AGN feedback, as AGN feedback moves gas from the centre of the haloes and redistributes it to the exterior. This also quenches star formation in galaxies, further lowering the clustering at small scales, seen as a suppression of the 2pCF and $P(k)$ on these scales. However, we do not observe the same amount of suppression in the total matter component that we observe in the gaseous component because a significant contribution to the total matter comes from the DM, which remains largely unaffected due to AGN feedback.

\begin{figure*}
    \centering
    \includegraphics[width=\textwidth]{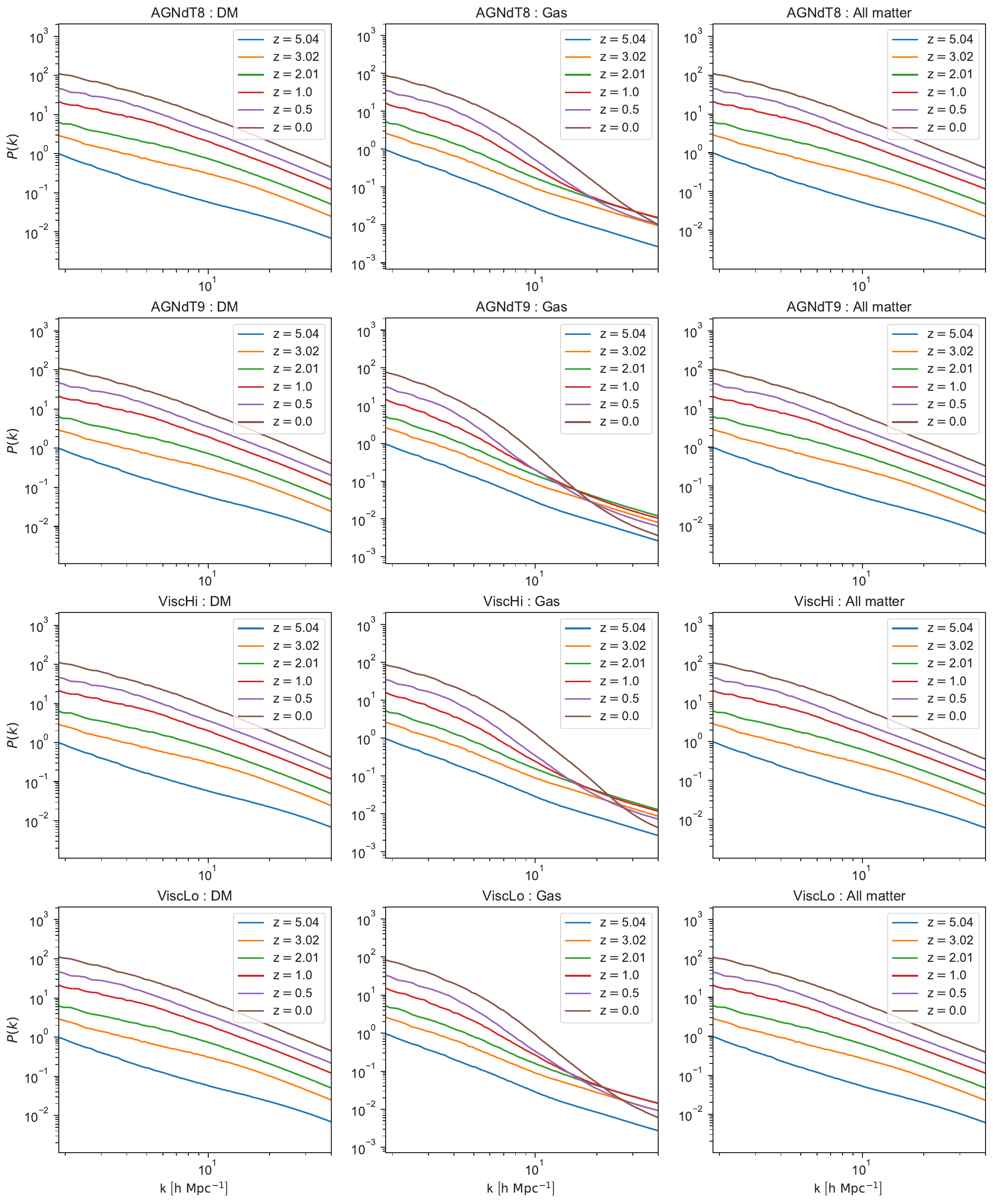}
    \caption{$P(k)$ for the different models that we studied at different epochs. The crossover events are evident in the gaseous component in all the models. As time goes by, matter starts clustering which boosts the $P(k)$ on all scales; however AGN feedback redistributes the gas, which suppresses the $P(k)$ at small scales, but boosts it at intermediate scales.}
    \label{fig:figA3}
\end{figure*}

\begin{figure*}
    \centering
    \includegraphics[width=\textwidth]{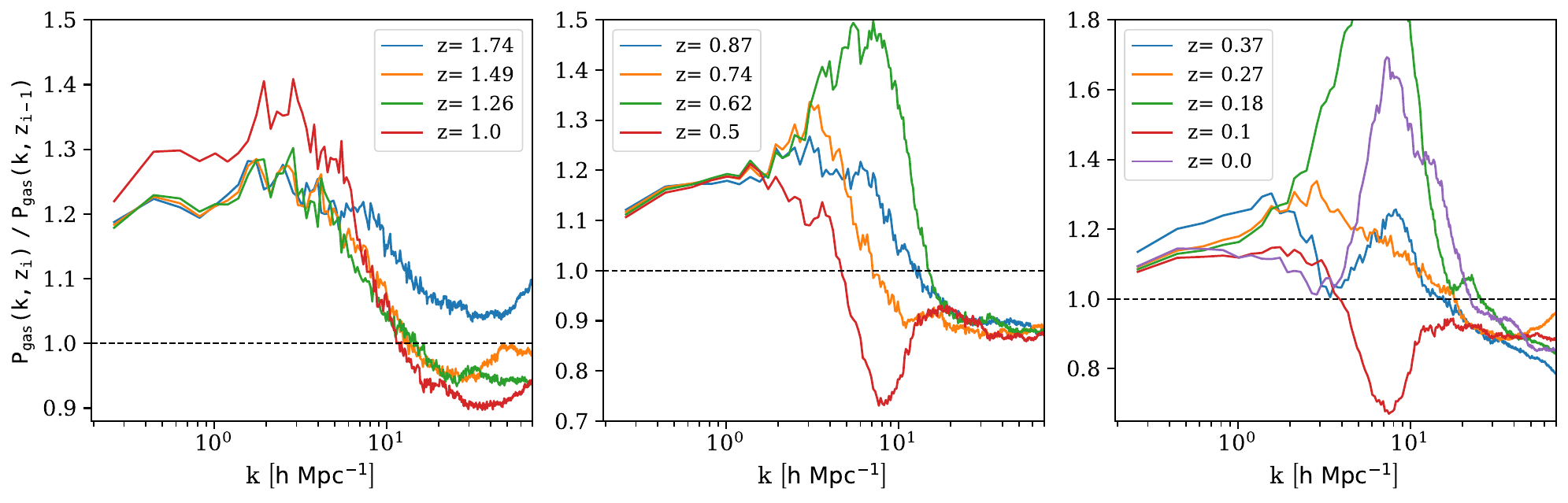}
    \caption{Ratio of power spectrum of gas between two consecutive epochs for {\tt AGNdT9} model. The panels are arranged in decreasing redshift if we go from left to right. The ratio significantly falls below 1 at $z=1.49$ (first panel from left) which marks the onset of dominance of AGN feedback.}
    \label{fig:figA4}
\end{figure*}

For completeness, in Figure~\ref{fig:figA3}, we also present the $P(k)$ of all the models we studied at different epochs. We observe crossover events discussed in Section~\ref{subsec:models} in the gaseous component in all models at $z\approx 1-1.5$, which agrees with our constraints on the dominance of AGN feedback from {\tt AGNdT8} and {\tt AGNdT9} models.

To determine the time scale at which AGN feedback starts to dominate, we plot in Figure \ref{fig:figA4} the ratio of $P(k)$ between two consecutive epochs for {\tt AGNdT9}, similar to the Figure \ref{fig:fig10}. We observe that, in this case, dominance of AGN feedback starts a bit earlier, at around $z=1.49 - 1.26$. This is due to the higher $\Delta T_{AGN}$ parameter for the black holes in {\tt AGNdT9} model that make the black holes more efficient in heating the surrounding and transportation of the gas due to AGN feedback. Also compared to the same plot for {\tt AGNdT8} in Figure \ref{fig:fig10}, the changes in {\tt AGNdT9} are much more drastic, particularly at late times, i.e $z=0.18 - 0$. We observe a stronger and rapid change in characteristics of the curve (i.e change in the slopes of the curve) in the {\tt AGNdT9} model.

\section{Effect of Box Size}
\begin{figure}
    \centering
    \includegraphics[width=\linewidth]{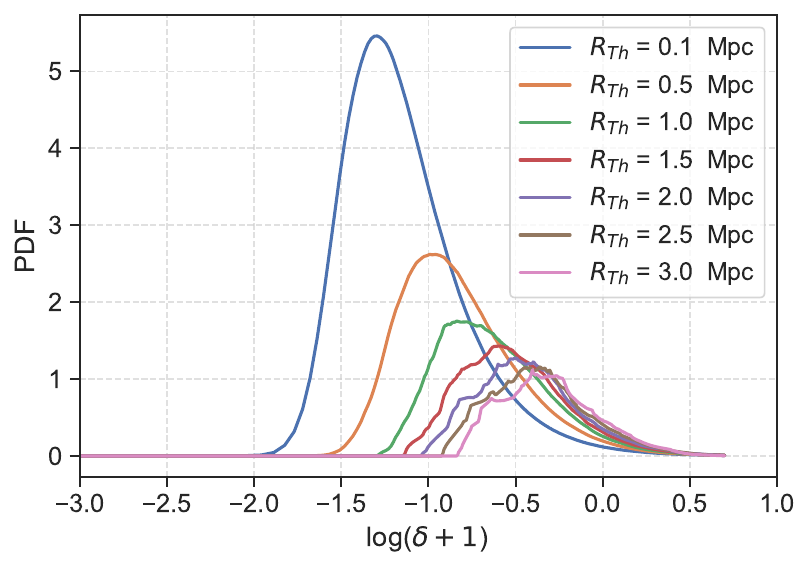}
    \caption{Distribution of gas at $z=0$ for {\tt AGNdT8}. The distribution is skewed. It starts to be a mean 0 distribution at very large smoothing scales }
    \label{fig:figB2}
\end{figure}

\begin{figure}
    \centering
    \includegraphics[width=\linewidth]{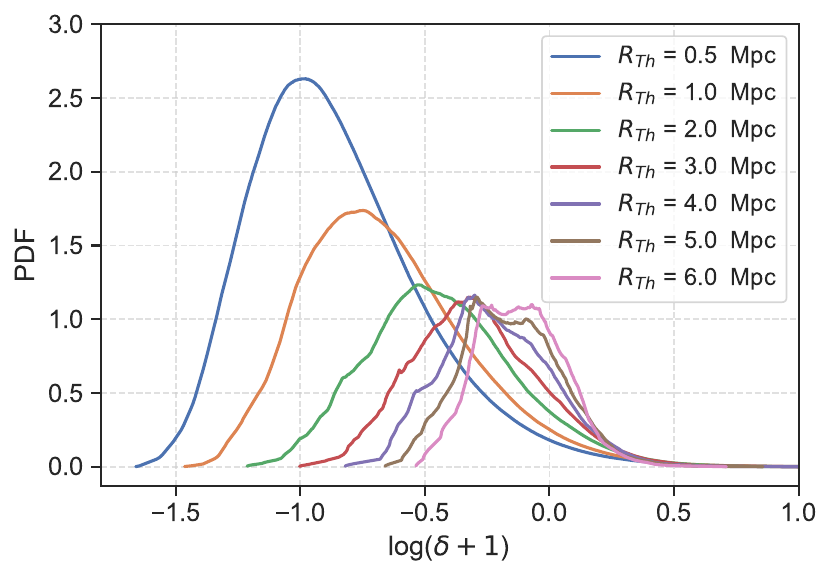}
    \caption{Distribution of gas at $z=0$ for the EAGLE Reference Model with Box Size of 100 Mpc. The distribution is skewed. It starts to be a mean 0 distribution at very large smoothing scales, and does not differ significantly from the {\tt AGNdT8} model.}
    \label{fig:figB3}
\end{figure}

\begin{figure}
    \centering
    \includegraphics[width=\linewidth]{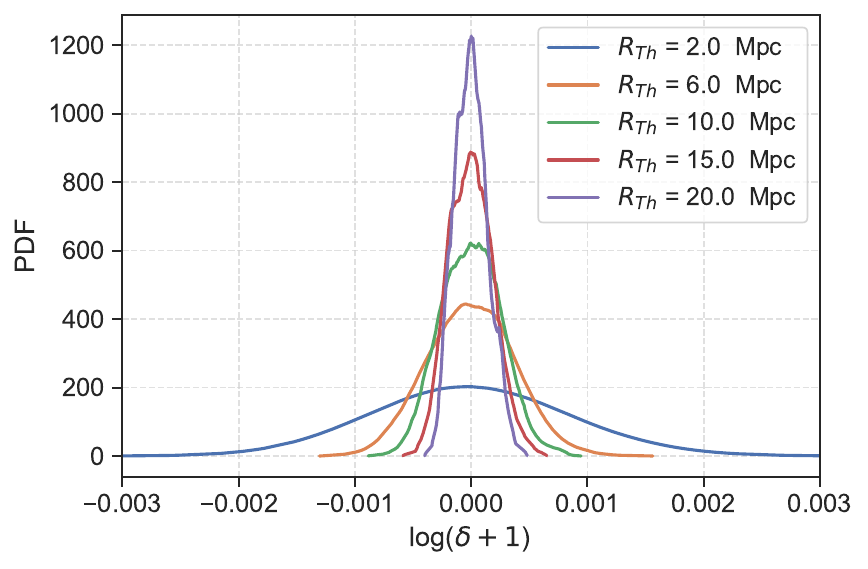}
    \caption{Distribution of gas at $z=0$ for TNG300-1. The distribution is a narrow Gaussian distribution with mean 0 which is expected for large box size. The width of the distribution decreases with increasing smoothing scale.}
    \label{fig:figB4}
\end{figure}

Cumulants are highly sensitive to the size and resolution of the box. To have a complete large-scale picture, we need a larger simulation box size without compromising resolution. To demonstrate how different box size and resolution can effect the statistical quantities, we plot the distribution of the first cumulant, i.e the PDF of the density field for different smoothing scales for  simulations with different box sizes. 

In Figure~\ref{fig:figB2}, we plot the distribution of gas at various smoothing scales for {\tt AGNdT8}, in Figure~\ref{fig:figB3} we plot the same quantities for the EAGLE reference model with a box size of $100$ Mpc. Furthermore, in Figure~\ref{fig:figB4}, we compare these distributions to the IllustrisTNG-300-1 simulation. The IllustrisTNG suite \citep{Pillepich_2017} is comprised of multiple runs with different box sizes and resolutions. We use the TNG300-1 run which has a periodic box size of $L=205~h^{-1}~\mathrm{Mpc} \approx 300~\mathrm{Mpc}$ on a side and uses $2\times2500^3$  resolution elements, and uses the Planck 2016 \citep{Planck_2016} cosmology.

We see that the box size affects the distribution because the distribution in {\tt AGNdT8} is a skewed Gaussian, whereas, for TNG300-1, it becomes a narrower and peaks around 0 as we increase the smoothing scale. We also note that the differences between the {\tt AGNdT8} model and the EAGLE reference model is not as significant as the differences between the {\tt AGNdT8} and the TNG300-1 model. This further strengthens the need to larger box sizes as the statistical properties can be significantly different for the larger box. This variation will be more prominent if we look at the higher order cumulants from TNG300-1. The larger box size in TNG300-1 results in a more uniform distribution, which will also lower the magnitude of the cumulants. However, we note that the qualitative comparison between {\tt AGNdT8}, {\tt AGNdT9} made throughout Section~\ref{sec:results} remains valid.

\section{Effect of sampling on 3pCF}
\label{sec:3pCF-Sampling}
In Section~\ref{sec:3pCF} we computed the 3pCF for the distribution of gas for {\tt AGNdT8} and {\tt AGNdT9}. For computing the 3pCF, we first take 20 random uniform samples from the entire dataset containing $0.1\%$ of the total gas particles; this is done to reduce computational cost. In this appendix, we consider the error induced by sub-sampling the gas distribution. 

\begin{figure}
    \centering
    \includegraphics[width=\linewidth]{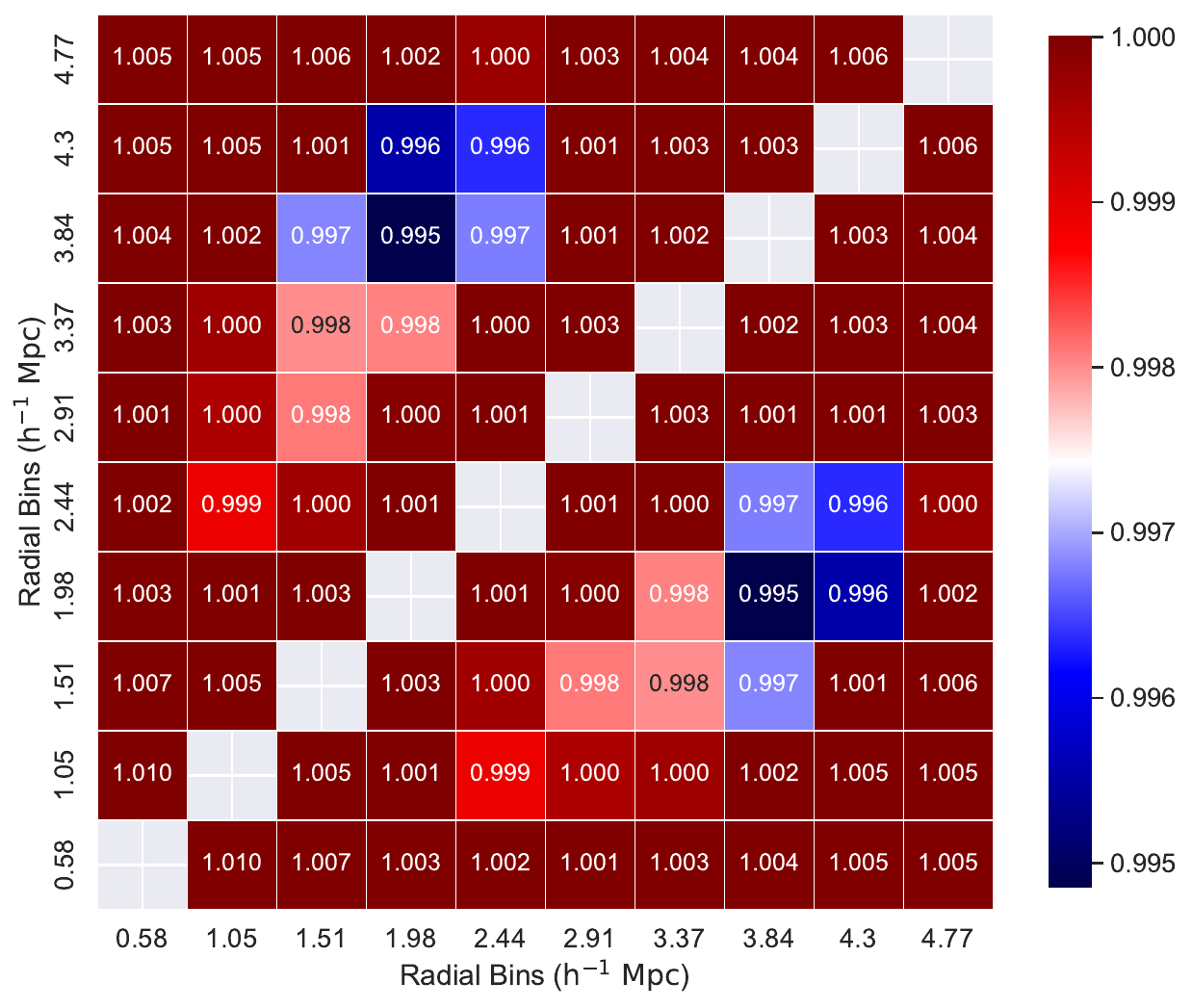}
    \caption{Ratio of 3pCF between $0.1\%$ and $0.05\%$ subset of the complete data}
    \label{fig:figC1}
\end{figure}

\begin{figure}
    \centering
    \includegraphics[width=\linewidth]{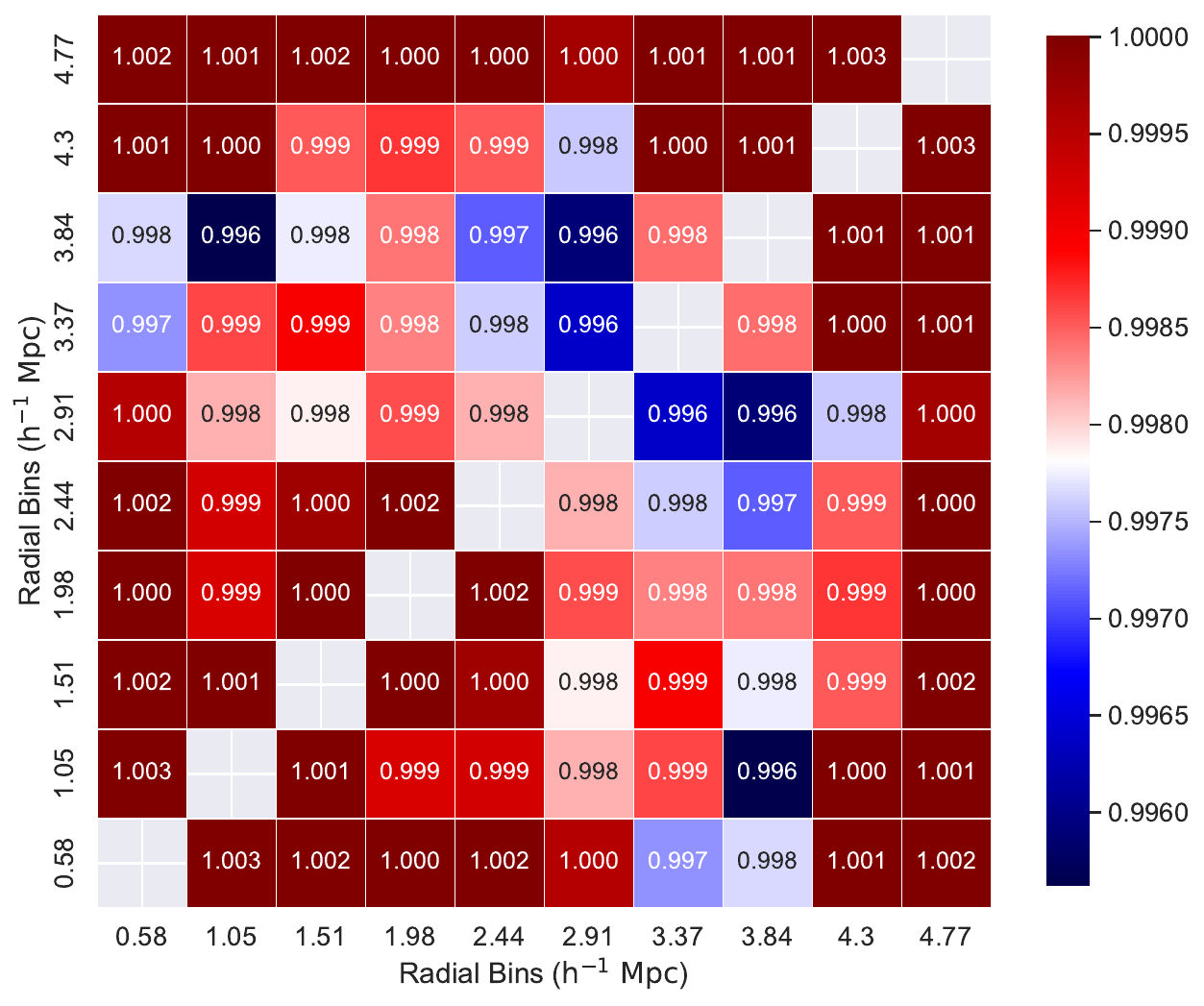}
    \caption{Ratio of 3pCF between $0.1\%$ and $0.5\%$ subset of the complete data}
    \label{fig:figC2}
\end{figure}

\begin{figure}
    \centering
    \includegraphics[width=\linewidth]{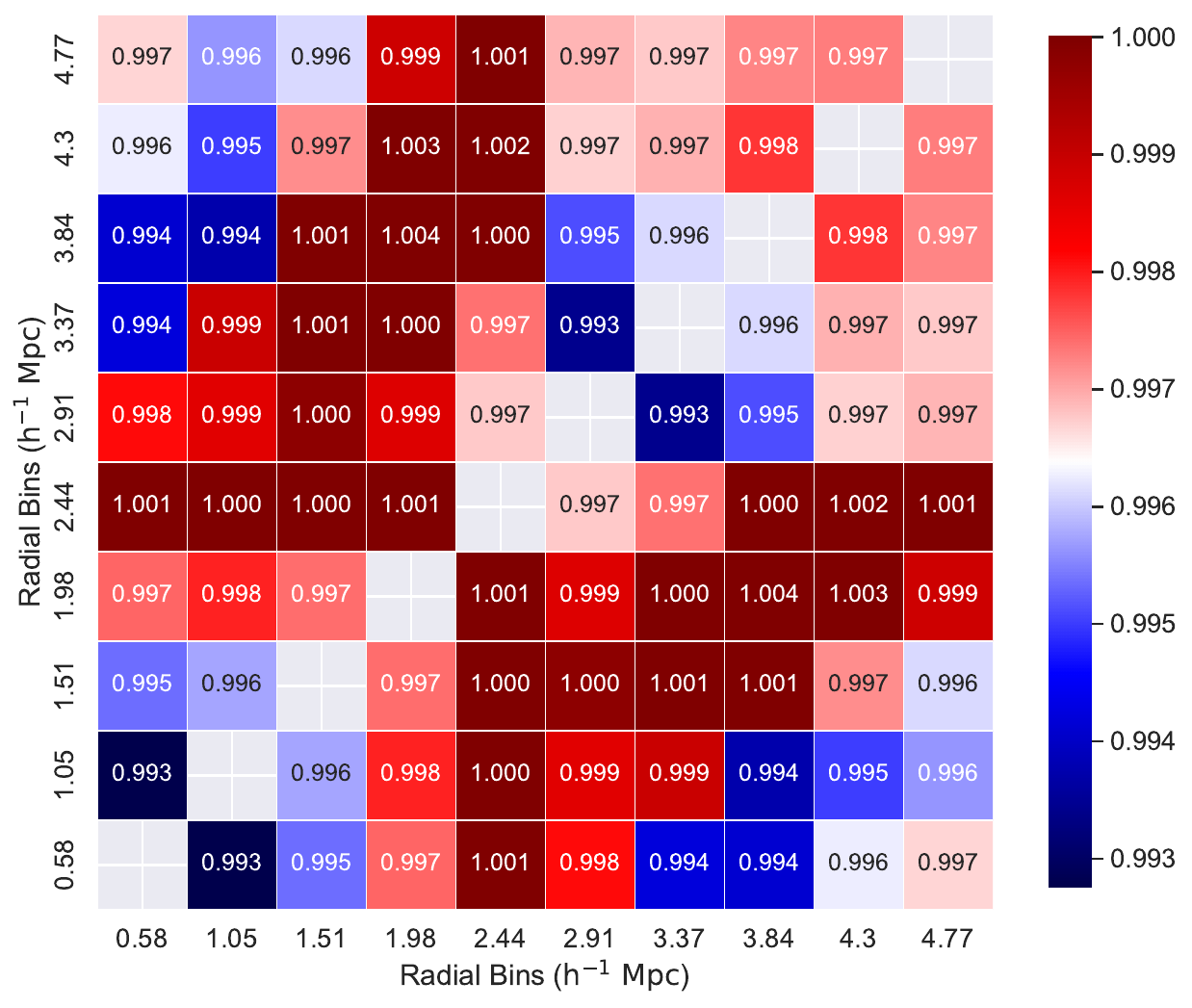}
    \caption{Ratio of 3pCF between $0.5\%$ and $0.05\%$ subset of the complete data}
    \label{fig:figC3}
\end{figure}

In Figure~\ref{fig:figC1} we compute the ratio of 3pCF of {\tt AGNdT9} with $0.1\%$ subset of the data to the 3pCF of {\tt AGNdT9} with $0.05\%$ sample of the complete dataset. Similarly in Figure~\ref{fig:figC2} we compute the ratio of 3pCF of {\tt AGNdT9} with $0.1\%$ subset of the data to the $0.5\%$ sample of the complete dataset, and in Figure~\ref{fig:figC3} we compute the ratio of 3pCF of {\tt AGNdT9} with $0.5\%$ subset of the data to the $0.05\%$ sample of the complete dataset. From these figures we can see that our choice of sampling the data does not affect the results significantly and the small deviation (on average $0.3\%$) is just due to random sampling.

The only effect of sampling is that, by taking a small subset of the data we are unable to retain the minute details of the matter distribution. However this issue less pronounced on larger scales which are sampled densely enough even in our $0.1\%$ subsamples to appreciably measure the 3pCF and provide a qualitative understanding of the matter distribution. 

\begin{figure}
    \centering
    \includegraphics[width=\linewidth]{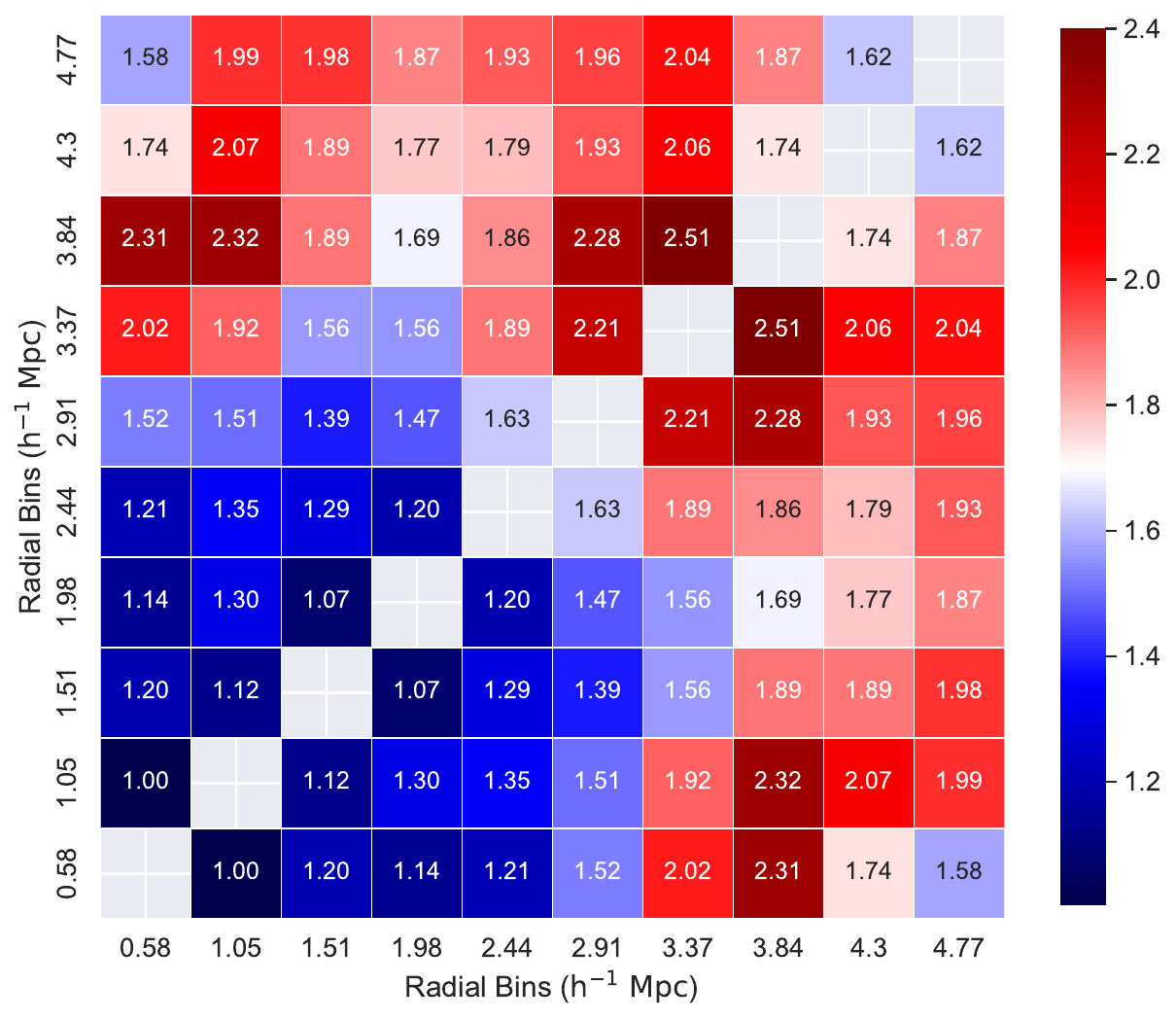}
    \caption{Ratio of 3pCF between {\tt AGNdT8} and TNG50-2. The 3pCF matrix shows small differences that vary in amplitude as a function of scale. This variation can be attributed to different resolutions of the two simulations but perhaps more likely due to different approaches for modelling AGN feedback. This shows the sensitivity of the 3pCF to variations in the subgrid model.}
    \label{fig:fig_C4}
\end{figure}

Finally, we also compare the 3pCF matrix for TNG50-2, the second highest resolution 50 Mpc box of the TNG suite. 
In Figure~\ref{fig:fig_C4}, we compare the 3pCF of {\tt AGNdT8} and TNG50-2. The magnitude of 3pCF of TNG50-2 is very comparable to the {\tt AGNdT8} model. The differences are attributed to different resolutions and AGN feedback models in the TNG Simulation. The TNG50-2 has a mass resolution of $m_{\rm gas} = 6.8\times 10^5 ~{\rm M_\odot}$
and uses $2\times1080^3$ particles, which is $2.5$ times the number of particles used in the EAGLE models considered here. The differences in the subgrid models are borne out in the the 3pCF -- where we notice that the ratio of the 3pCF between {\tt AGNdT8} to TNG50-2 is larger than that between {\tt AGNdT8} and {\tt AGNdT9}. There is a also a modest gradient to note in Figure~\ref{fig:fig_C4}, where differences increase somewhat as a function of separation. The structure of the 3pCF matrix therefore captures some of the scale-dependent differences between different models of AGN feedback, at least as modelled in TNG and EAGLE. 

\bsp	
\label{lastpage}
\end{document}